\documentclass[9pt,shortpaper,twoside,web]{ieeecolor}%短文9pt,shortpaper 长文journal
\usepackage{generic}
\usepackage{amsmath,amssymb,amsfonts}
\usepackage{algorithmic}
\usepackage{graphicx}
\usepackage{algorithm,algorithmic}
\usepackage{textcomp}
\usepackage{bm}
\usepackage{subfigure}
\usepackage[colorlinks=true,linkcolor=blue,citecolor=blue]{hyperref}
\allowdisplaybreaks
\def\BibTeX{{\rm B\kern-.05em{\sc i\kern-.025em b}\kern-.08em
    T\kern-.1667em\lower.7ex\hbox{E}\kern-.125emX}}
\begin{document}
\title{%Modified Prescribed Performance Control of High-Order MIMO Uncertain Nonlinear Systems with Input Saturation: A Perturbed Perspective
%Low-Complexity Distributed Control of High-Order MIMO Uncertain Nonlinear Systems with Input and Performance Constraints
Low-Complexity Control Under Input Saturation and   Performance Constraints: A Bidirectional Modification Scheme
%Low-Complexity Control for MIMO Systems with Input Saturation and Performance Constraints: A Bidirectional Modification Approach
%Tracking Control under Input Saturation and Performance Constraints: A Bidirectional Modification Approach
%Input-Saturated Tracking Control for MIMO Nonlinear Systems with Performance Constraints: A Bidirectional Modification Approach
}
\author{Jinyu Ni, Xiucai Huang, George A. Rovithakis, \IEEEmembership{Senior Member, IEEE}, and Yamin Yan, \IEEEmembership{Member, IEEE}
\thanks{%This work was supported in part by  the Fundamental Research Funds for the Central Universities under Grant No. 2024CDJYXTD-007; in part by the Natural Science Foundation of Chongqing under Grant No.CSTB2023NSCQ-LZX0026; in part by the Project of New Chongqing Youth Innovation Talent under Grant No. 2024NSCQ-qncxX0433; in part by the National Natural Science Foundation of China under Grant No. W2411061, and Grant No. 62303080; in part by the Natural Science Foundation of Chongqing under Grant No. CSTB2022NSCQ-MSX0609; in part by the Chongqing Human Resources and Social Security Bureau  under Grant No. cx2021114; and in part by the China Postdoctoral Science Foundation under Grant No. 2022M720570. \textit{(Corresponding author:  Xiucai Huang.)}
} 
\thanks{Jinyu Ni and Xiucai Huang are with the  International Joint Laboratory on Safety and Control of Autonomous Unmanned Systems of Ministry of Education, Chongqing University, Chongqing 400044, China (e-mail: jinyuni@foxmail.com; hxiucai@cqu.edu.cn%, zhaokai@cqu.edu.cn,  and ydsong@cqu.edu.cn
).}
\thanks{George A. Rovithakis is with the Department of Electrical and Computer Engineering, Aristotle University of Thessaloniki, Thessaloniki 54124, Greece (e-mail: rovithak@ece.auth.gr).}
\thanks{Yamin Yan is with the School of Electrical and Electronic Engineering, Nanyang Technological University, Singapore 639798 (e-mail: yamin.yan@ntu.edu.sg).}
}

\maketitle

\begin{abstract}
	This article addresses the output tracking control problem for a class of  high-order uncertain highly-coupled MIMO nonlinear systems subject to input saturation and performance constraints. To resolve the problem, a bidirectional modification mechanism is constructed, which is able to not only relax the constraints when saturation occurs to alleviate potential conflict, but also accelerate the recovery of original constraints after saturation ceases, and further tighten the constraints to enhance control performance if saturation remains inactive at the steady-state phase. Based on the mechanism, a model-, approximation- and complexity-explosion-free control scheme is proposed. To bypass the obstacle in Lyapunov analysis, a novel stability analysis framework is developed, which, given that two parameter selection conditions are met, ensures satisfaction of modified constraints and boundedness of all closed-loop signals. Simulation results validate the effectiveness and superiority of the methodology.
\end{abstract}


\section{Introduction}
\label{sec:introduction}
%\subsection{Issues of Prescribed Performance Control}
%\subsection{Input-Saturated Prescribed Performance Control}
%\IEEEPARstart{T}{o} 
\textcolor{black}{To reliably fulfill various tasks, guaranteeing satisfaction of performance constraints (i.e., enforcing specifications of overshoot, convergence speed and steady-state accuracy on the output error)  has long been a central objective in the control of real-world systems \cite{BIKAS2026,NI2025112332}.} % \cite{9351643}. 
Simultaneously, to facilitate  practical implementation, it is also highly desirable  that the controller can address inevitable constraints of input saturation  \cite{11428271}, %\cite{9985424,11428271}, 
while maintaining low structural and computational complexity, i.e., avoiding exact model knowledge, approximation tools like neural networks (NN) or fuzzy logic systems (FLS), and complexity-explosion-issue caused by backstepping design \cite{11205324}. %This entails avoiding reliance on exact model knowledge, approximation tools such as neural networks (NNs) or fuzzy logic systems (FLSs), and the complexity explosion issue caused by backstepping design \cite{NI2025112332}.
%the structural and computational complexity of  controller remains at a low level. This entails avoiding reliance on exact model knowledge, approximation tools such as neural networks (NNs) or fuzzy logic systems (FLSs), %adaptive techniques, 
%and complexity explosion issue caused by standard backstepping design  \cite{10101860}.

Prescribed Performance Control (PPC) is a %representative
typical low-complexity solution for imposing performance constraints on uncertain nonlinear systems \cite{BECHLIOULIS20141217}. %  \cite{BECHLIOULIS20141217,11205324,NI2025112332}. %and the references therein 
In PPC design, an appropriate performance function  is first selected according to the desired specifications. %Subsequently,
Then, the output error is normalized by the function and mapped by a barrier function, thus transforming the constrained system into an unconstrained one. Finally, the latter is proven strictly stable, thus ensuring satisfaction of performance constraints. 
%By selecting appropriate performance functions, PPC can guarantee that the desired performance specifications are met as prescribed by the user. 
However, %to satisfy these constraints, 
PPC often requires relatively large control effort, particularly when the specifications are stringent, which may render it invalid under input saturation. %inevitably impedes its  practical deployment since the physical capability of actuators are inherently restrictive. 
While substantial efforts have been devoted  to solving the issue,
many results rely on multiple adaptive laws \cite{mei2023ilc}, NN \cite{cao2023neuroadaptive}, FLS \cite{yin2024fuzzy}, observers \cite{feng2019novel} or backstepping design  \cite{song2021prescribed} to cope with uncertainties,
%and the accordingly introduced components (e.g., multiple adaptive laws \cite{zhao2023novel,mei2023ilc}, %\cite{zhao2023novel,mei2023ilc,shen2018performance,lyu2021predefined}, 
%NNs \cite{chang2022adaptive,cao2023neuroadaptive}, %\cite{chang2022adaptive,xia2023prescribed,mei2023ilc,cao2023neuroadaptive}, 
%FLSs \cite{yin2024fuzzy}, %disturbance \cite{feng2019novel} or extended state \cite{yang2018prescribed}   observers
%or uncertainty observers \cite{feng2019novel,yang2018prescribed}) and backstepping design  \cite{xia2023prescribed,shen2018performance,lyu2021predefined,shao2018fault,song2021prescribed}
% \cite{chang2022adaptive,xia2023prescribed,mei2023ilc,yin2024fuzzy,feng2019novel,shen2018performance,yang2018prescribed,lyu2021predefined,shao2018fault,song2021prescribed}
which significantly increases the controller complexity. %In \cite{11234908}, the  compensation term employs the signum function, which could induce chattering. 
Moreover, a critical issue is often overlooked, i.e., if the control task is feasible under input saturation. For example, the system $\dot x=2+\operatorname{sat}(u)$ %where $x$ is the state and $u$ is the input
with saturation level $\bar u=1$ %cannot be stabilized 
is unstable as $\dot x$ is always positive. Hence, the problem to preserve low-complexity of controller while ensuring feasibility of task arises.
	
%The \textit{third} %is the \textit{dynamic modification} strategy, which 
%gives an answer to the problem by establishing sufficient feasibility conditions on the saturation level, 
The constraint modification strategy gives an answer to the problem by establishing proper feasibility conditions on $\bar u$ \textcolor{black}{or assuming the plants are input-to-state practically stable (ISpS),} %where the constraints are relaxed when  saturation occurs and tightened to restore performance after saturation ceases, 
without involving high-complexity components. %, representing the latest development trend of input-saturated PPC.  
However, high-order \cite{ji2021saturation,ji2023saturation,11304721} or multiple first-order \cite{11130930,trakas2024adaptive} dynamics are needed to modify each constraint,  making controller complexity still high and compromises the design flexibility.  Furthermore, the feasibility condition requires $\bar u$ to be sufficiently large  \cite{11304721,11130930,gkesoulis2025low,10138653}, which incurs prohibitive hardware costs and is unrealistic %for many real-world systems
in practice.  More fundamentally, the constraints can only be relaxed or restored passively, rendering the overall control performance conservative. %when saturation persists at the steady-state, the constraints in these methods remain relaxed, rendering the steady-state performance inherently conservative. %the relaxation of constraints during saturation could severely degrade control performance, rendering these results relatively conservative. %evolve along the single path of relaxation-recovery,  %adjust the constraints unidirectionally, %i.e., relax the boundary when saturation occurs and gradually recover to the nominal value after saturation disappears, 
%thus unable to alleviate performance degradation during saturation.
Other attempts %have also been made, e.g., 
using controller switching \cite{bikas2024prescribed}, sliding mode \cite{10004950}, or high-order reference modification \cite{10273607} are also reported. Yet,
%Some other answers are given via discontinuous controller switching \cite{bikas2024prescribed}, sliding mode control \cite{10004950}, or high-order reference modification \cite{10273607}. However, %discontinuous controller \cite{bikas2024prescribed}, conservative range for initial condition of performance function,  and 
$\bar u$ still should be large.

On the other hand, the class of plants in existing results remains relatively narrow, confined to SISO  \cite{10138653,bikas2024prescribed,10004950} or MIMO \cite{cao2023neuroadaptive} systems with state- or time-independent nonlinearities. While state- and time-dependent ones are involved in \cite{trakas2024adaptive,gkesoulis2025low,10273607}, the plants are limited to SISO \cite{gkesoulis2025low}, MIMO normal-form  \cite{trakas2024adaptive}, or strict-feedback systems with decoupled inputs \cite{10273607}. %with diagonal control gain matrices \cite{10273607}. 
Simultaneously, most of these results are built upon conservative assumptions, e.g., system nonlinearities or external disturbances are bounded \cite{11304721}, periodic \cite{mei2023ilc}, known \cite{yin2024fuzzy,feng2019novel,ji2023saturation},  %or positive definite (gain matrices) \cite{cao2023neuroadaptive}, %, or tailored \cite{bikas2024prescribed}, 
which may be difficult to validate and seldom satisfied in practice, thereby restricting the applicability of these methods. Moreover, high-order derivatives of the desired trajectory are required in \cite{trakas2024adaptive}, %\cite{shen2018performance,cao2023neuroadaptive,trakas2024adaptive}, 
rendering the controller sensitive to measurement noise.

In light of the preceding background, we present  a continuous, low-complexity, bidirectional modification  input-saturated PPC solution in this article. The major contributions of our work %beyond the state of the art 
are as follows.
%\begin{itemize} 
%\item 
	%\textit{Construction of a bidirectional constraint modification system.} %To bias the trade-off between input saturation and performance constraints  toward the latter, 
	%Since a trade-off between input saturation and performance constraints is inevitable, 
	%Since the control performance could be sacrificed when input saturation occurs, 
	%\textcolor{red}{To reduce the degradation of control performance   during input saturation,} 
\textit{First}, to reduce the conservatism of control performance in existing results, we construct a targeted \textit{bidirectional modification mechanism}, %by exploiting the interplay between the saturation deficiency and a performance enhancement term. 
	which is able to not only relax  the constraints when saturation occurs to alleviate potential conflict, but also accelerate the recovery of original constraints after saturation ceases, and further tighten the constraints to enhance performance if saturation remains inactive at the steady-state phase.  %reduces the relaxation of 
	 %as little as possible 
	 %it accelerates their tightening. 
	%Notably, the control performance can be further enhanced if saturation is mostly inactive.
	%mitigates input saturation while alleviates performance degradation  as much as possible, and can even further enhance control performance when saturation is generally mild.
	Moreover, %unlike the results  employing high-order  \cite{ji2021saturation,ji2023saturation,11304721,10273607} or multiple first-order \cite{11130930,trakas2024adaptive} dynamics, 
	our modification dynamics is both first-order and single, %whose output directly adjusts all  constraints, 
	thereby significantly reducing the  complexity introduced into controller and increasing the flexibility of design.
	%\item %\textit{Development of a new design and stability analysis framework.} %To support the construction and generalization described above, novel design elements and a tailored stability analysis are required. 
	\textit{Second}, based on the  mechanism, %and deliberate weighting of the modification signal, 
	we design a \textit{continuous and low-complexity control scheme}, neither imposing conditions on %relaxing the %need for prior knowledge of the saturation level as in \cite{11234908,zhao2023novel,chang2022adaptive,xia2023prescribed,yang2018prescribed,shao2018fault,cao2023neuroadaptive,song2021prescribed,https://doi.org/10.1002/rnc.70022}, 
	%reliance of feasibility condition on 
	$\bar u$ %as in \cite{11304721,gkesoulis2025low,11130930,10138653,bikas2024prescribed,10004950,10273607} 
	nor requiring high-order derivatives of the desired trajectory. %as in \cite{shen2018performance,cao2023neuroadaptive,trakas2024adaptive}. 
	To bypass the obstacle in Lyapunov analysis (Remark 6), we develop a  \textit{novel stability analysis framework}. %\textcolor{black}{by combining proof by contradiction, BIBS stability and barrier functions.} 
	Given that two  parameter selection conditions  are met, it rigorously establishes satisfaction of  modified constraints and boundedness of all closed-loop signals. Furthermore, %contrary to \cite{11130930,10138653,ji2021saturation,ji2023saturation,bikas2024prescribed} where guidance to adjust  the modified performance is lacking, %(i.e., no explicit parameter selection criterion is provided), 
	we present in Theorem 2 a clear parameter adjustment criterion, which characterizes the dependence of modification signal's bounds on parameters, offering a route to manage the trade-off between modification magnitude and allowable tightening of constraints. %The proposed control scheme does not rely on prior knowledge of the saturation level or on any feasibility condition tied to its lower bound, thereby surpassing the limitations of \cite{11234908,zhao2023novel,chang2022adaptive,xia2023prescribed,yang2018prescribed,shao2018fault,cao2023neuroadaptive,song2021prescribed,gkesoulis2025low,11130930,10138653,11304721,https://doi.org/10.1002/rnc.70022}. requiring high-order derivatives of the desired trajectory as in \cite{shen2018performance,cao2023neuroadaptive,trakas2024adaptive}
	%maximum achievable performance enhancement.
	%\item %\textit{Generalization of the  controlled plant class.}
	\textit{Third}, supported by the  framework,  the results are extended to a broader class of \textit{high-order uncertain highly-coupled MIMO strict-feedback nonlinear systems}, where  %high-order strict-feedback MIMO uncertain nonlinear systems. 
	%Superior to \cite{11234908,zhao2023novel,chang2022adaptive,yin2024fuzzy,feng2019novel,yang2018prescribed,shao2018fault,cao2023neuroadaptive,song2021prescribed, gkesoulis2025low,10138653,trakas2024adaptive,https://doi.org/10.1002/rnc.70022}, %\cite{11234908,zhao2023novel,chang2022adaptive,yin2024fuzzy,feng2019novel, shen2018performance,yang2018prescribed,lyu2021predefined,shao2018fault,cao2023neuroadaptive,song2021prescribed, gkesoulis2025low,10138653,trakas2024adaptive,https://doi.org/10.1002/rnc.70022}, 
	 the nonlinearities are both state- and time-dependent, without imposing  restrictive assumptions. % as in \cite{zhao2023novel,chang2022adaptive,mei2023ilc,yin2024fuzzy,feng2019novel,cao2023neuroadaptive,ji2023saturation,11304721,https://doi.org/10.1002/rnc.70022,10004950}. %; and more importantly, relaxing the need to impose controllability condition on the  gain matrices as in \cite{trakas2024adaptive}. 
%\end{itemize}
%\textit{Added Value Relative to the Conference Version \textcolor{black}{[31]}:} %\textcolor{black}{Compared with its conference version []}, 

\textcolor{black}{The \textit{added value} of this paper relative to its conference version \cite{Ni2026CSIS} includes: 1) a generalization from second-order SISO systems to higher-order MIMO systems; 2) updated and more detailed analysis, designs, and proofs; 3) the inclusion of Theorem~2, which guides for parameter adjustment; and 4) a comparative simulation. % an numerical example, a comparative simulation, on a mass-spring-damper system, and a parameter sensitivity analysis. %Taken together, these additions and refinements 
These contribute to completeness of theory and reliability of verification.} %an additional \textcolor{black}{nine} pages. %beyond the conference version \textcolor{black}{[31], Ar-xiv}. 

\textcolor{black}{\textit{Notation:}} %$\mathbb N$ denotes the set of natural numbers. 
%$\mathbb R$ is the set of real numbers. $\mathbb R_{\ge 0}$ is the set of non-negative real numbers. 
$\mathbb R$ ($\mathbb R_{\ge 0}$) is the set of (non-negative) real numbers. Given integers $a,b>0$, $\mathbb R^a$ is the $a$-dimensional Euclidean space, $\mathbb R^{a\times b}$ is the set of $a\times b$ real matrices, and $\mathbb{I}_a^b$ is the set of integers from $a$ to $b$. $\|\diamond\|_1$ and $\|\diamond\|$ denote the $\mathcal{L}_1$ and $\mathcal{L}_2$ norms of vector $\diamond$, respectively. $\operatorname{diag}(\diamond_1,...,\diamond_a)$ is a diagonal matrix with entries $\diamond_1$-$\diamond_a$. %$I_a$ denotes the $a\times a$ identity matrix. For vectors $v=[ v_{1},...,v_{m} ]^\top$ and $w=[  w_1,...,w_m ]^\top$, the inequalities $v\ge w$ and $v>w$ are understood component-wise, i.e., $v_j\ge w_j$ and $v_j> w_j$ for all $j\in\mathbb I_1^m$, respectively. Given a matrix $M=[ M_{j,l} ]$ with $l\in\mathbb I_1^m$, $M\ge 0$ and $M> 0$ indicate that $M_{j,l}\ge 0$ and $M_{j,l}> 0$, respectively. For brevity, arguments of some functions are omitted or replaced by $(\diamond)$.

\section{%Preliminaries and 
	Problem Statement}
\label{sec:problem}

\subsection{System Description}
Consider the following MIMO nonlinear systems ($i\in\mathbb I_1^{n-1}$):
%Let us consider the following class of high-order  uncertain highly-coupled MIMO nonlinear systems ($i\in\mathbb I_1^{n-1}$): %\textcolor{black}{up-actuated systems}
\begin{flalign}
			\dot x_{i}(t)=&\ f_{i}(\tilde x_{i}(t),t%d_i(t)
			)+g_{i}(\tilde x_{i}(t),t)x_{i+1}(t)%, \ \  i\in\mathbb I_1^{n-1}
		\notag\\
		\dot x_{n}(t)=&\ f_{n}(\tilde x_{n}(t),t)+g_{n}(\tilde x_{n}(t),t)\operatorname{sat}(u(t))\notag\\
		y(t)=&\ x_{1}(t)
	 \label{sys}
\end{flalign}
where  $x_{i}(t)=[ x_{i,1}(t),...,x_{i,m}(t)  ]^\top\in \mathbb  R^m$, $i\in\mathbb I_1^{n}$ are system states, $\tilde x_{i}(t):=[ x_{1}^\top(t),...,x_{i}^\top(t) ]^\top\in \mathbb R^{mi}$, %$d_i(t)=[ d_{i,1}(t),...,d_{i,m_{d_i}}(t)  ]^\top\in \mathbb R^{m_{d_i}}$ are bounded and piecewise continuous unknown disturbance signals, 
$y(t)=[ y_{1}(t),...,y_{m}(t)  ]^\top\in \mathbb  R^m$ are output variables,  $u(t)=[  u_{1}(t),...,u_{m}(t) ]^\top\in \mathbb R^m$ are control inputs, and $\operatorname{sat}(u(t)):=[ \operatorname{sat}(u_{1}(t)),...,\operatorname{sat}(u_{m}(t)) ]^\top\in \mathbb R^m$ are saturated control inputs with $\operatorname{sat}(u_{j}(t))$, $j\in\mathbb I_1^m$  being defined as
\begin{flalign}
	\operatorname{sat}(u_{j}(t)):=\left\{ \begin{array}{r}
		u_{j}(t),\ \ \text{if}\ |u_{j}(t)|< \bar u_{j}\\
		\bar u_{j} \operatorname{sgn}(u_{j}(t)),\ \ \text{if}\ |u_{j}(t)|\ge \bar u_{j}\\
	\end{array} \right. \label{sat}
\end{flalign}
where $\bar u_{j}>0$ is the saturation level of control input $u_{j}(t)$, and $\operatorname{sgn}(\diamond)$ is the signum function. \iffalse defined as
\begin{flalign}
	\operatorname{sgn}(\diamond):=\left\{ \begin{array}{r}
		-1,\ \ \text{if}\ \diamond< 0\\
		0,\ \ \text{if}\ \diamond= 0\\
		1,\ \ \text{if}\ \diamond> 0\\
	\end{array} \right. .
\end{flalign}\fi 
In addition, $f_{i}:\mathbb R^{mi}\times \mathbb %R^{m_{d_i}}
R_{\ge 0}\rightarrow \mathbb R^m$ with $f_{i}(\tilde x_{i}(t),t)=[ f_{i,1}(\tilde x_{i}(t),t),...,f_{i,m}(\tilde x_{i}(t),t) ]^\top$ and $g_{i}:\mathbb R^{mi}\times \mathbb R_{\ge 0}\rightarrow \mathbb R^{m\times m}$ with $g_{i}(\tilde x_{i}(t),t)=[ g_{i,j,l}(\tilde x_{i}(t),t) ]$, $l\in\mathbb I_1^m$, respectively, are unknown vector and matrix functions, which are locally Lipschitz in $\tilde x_{i}(t)$ and piecewise continuous in $t$.  \iffalse Particularly, $f_{i}(\tilde x_{i}(t),t)$ and $g_{i}(\tilde x_{i}(t),t)$ are denoted by  %$f_{i}(\tilde x_{i},t)=[ f_{i,1}(\tilde x_{i},t),...,f_{i,m}(\tilde x_{i},t) ]^\top$ and $g_{i}(\tilde x_{i},t)=[ g_{i,j,l}(\tilde x_{i},t) ]$, $l\in\mathbb I_1^m$.
\begin{flalign}
	f_{i}(\tilde x_{i}(t),t)&= \begin{bmatrix}
		f_{i,1}(\tilde x_{i}(t),t)\\
		\vdots\\
		f_{i,m}(\tilde x_{i}(t),t)\\
	\end{bmatrix}\\
	g_{i}(\tilde x_{i}(t),t)&= \begin{bmatrix}
		g_{i,1,1}(\tilde x_{i}(t),t)& \cdots& g_{i,1,m}(\tilde x_{i}(t),t)\\
		\vdots& \vdots& \vdots\\
		g_{i,m,1}(\tilde x_{i}(t),t)& \cdots& g_{i,m,m}(\tilde x_{i}(t),t)\\
	\end{bmatrix}.\notag
\end{flalign}\fi 

%Before we formulate the control problem, two assumptions are introduced as follows.
\iffalse
\textcolor{black}{\textit{Definition 1 (see \cite{bacciotti2000necessary}):} %(Bounded-input bounded-state stable, see \cite{bacciotti2000necessary}): (see \cite{bacciotti2000necessary}):}  
The system \(\dot x(t)=f(x(t),u(t),t)\)
 where $f: \mathbb R^{m_x}\times\mathbb R^{m_u}\times\mathbb R_{\ge 0}\to\mathbb R^{m_x}$ is said to be \textit{bounded-input bounded-state (BIBS) stable} if for any initial state $x(0)$ and any bounded input $u(t)$, its solution $x(t)$ remains bounded, $\forall t\ge 0$.}
\fi 

\textcolor{black}{\textit{Assumption 1:}} (\ref{sys})  is \textit{bounded-input bounded-state (BIBS)} stable. %, $\forall t\ge 0$. %bounded input bounded state stable, i.e., 

The definition of BIBS stable can be found in \cite{bacciotti2000necessary}.

\textcolor{black}{\textit{Assumption 2:}} For all $\tilde x_i(t)\in\mathbb R^{mi}$ and $t\ge 0$, there exist   \textcolor{black}{continuous} but not necessarily known functions $\bar f_i:\mathbb R^{mi} \rightarrow \mathbb R$ and $\bar g_i:\mathbb R^{mi} \rightarrow \mathbb R$, such that
\begin{flalign}
	\|f_i(\tilde x_i(t),t)\|\le|\bar f_i(\tilde x_i(t))|,\ \ %,\ \ \forall \tilde x_i(t)\in\mathbb R^{mi},\ \ \forall t\ge 0 \notag\\
	\|g_i(\tilde x_i(t),t)\|\le|\bar g_i(\tilde x_i(t))|. %,\ \ \forall \tilde x_i(t)\in\mathbb R^{mi},\ \ \forall t\ge 0.
\end{flalign}

%\textcolor{black}{\textit{Assumption 2:}} The symmetric matrices $g_i^\top(\tilde x_i,t)+g_i(\tilde x_i,t)$, $i\in\mathbb I_1^{n-1}$ are either positive definite or negative definite. No loss of generality, we assume that $g_i^\top(\tilde x_i,t)+g_i(\tilde x_i,t)$ are positive definite.

\textcolor{black}{\textit{Remark 1:}} \textcolor{black}{Assumption 1 can also be found in the standard literature \cite{trakas2024adaptive,10273607,5723705}, which is reasonably imposed to exclude plants that become unstable by input saturation.} As noted in Section \ref{sec:introduction}, the system $\dot x=2+\operatorname{sat}(u)$ with $\bar u=1$ is unstable and is therefore ruled out by Assumption 1. \textcolor{black}{A common practical example satisfying Assumption 1 is the mass-spring-damper system \cite{10273607}, which is widely employed in vehicle suspensions, seismic base isolators, precision manufacturing equipment, and aerospace structures.}
\textcolor{black}{For plants (\ref{sys}) that are non-BIBS stable, feasibility conditions on $\bar u_j$ need to be established, which is interesting yet challenging due to high coupling in the input channels, and deserves  further exploration.} Assumption 2 ensures that the nonlinearities cannot grow unbounded as $t$ increases.  %To ensure the controllability of system (\ref{sys}), it is commonly assumed  that  $g_i^\top(\tilde x_i(t),t)\Gamma_i(\tilde x_{i-1}(t),t)+\Gamma_i(\tilde x_{i-1}(t),t)g_i(\tilde x_i(t),t)$ are positive definite in the literature (e.g., \cite{10361274}), where the symmetric positive definite matrices $\Gamma_i(\tilde x_{i-1}(t),t)\in\mathbb R^{m\times m}$ may be chosen as diagonal \cite{NI2025112332} or the identity matrix \cite{trakas2024adaptive}. However, %no extra controllability condition is imposed on $g_i(\tilde x_i(t),t)$ such assumption is removed here, which thus further enlarges the scope of input-saturated PPC for MIMO systems in the form of (\ref{sys}).

\subsection{Control Objective}
\label{PF}
%In this article, we focus on the output tracking control problem for system (\ref{sys}) under both input saturation (\ref{sat}) and  performance constraints. 
Let $y_{d}:\mathbb R_{\ge0}\rightarrow\mathbb R^m$ with $y_d(t)=[  y_{d,1}(t),...,y_{d,m}(t) ]^\top$ be the vector of desired tracking trajectories, and let  $e(t)=[  e_{1}(t),...,e_{m}(t) ]^\top\in\mathbb R^m$ be the vector of output tracking errors with $e_j(t)$, $j\in\mathbb I_1^m$ being defined as
\begin{flalign}
	e_{j}(t):=x_{1,j}(t)-y_{d,j}(t).
\end{flalign}
Let $\rho_{1}:\mathbb R_{\ge0}\rightarrow\mathbb R^m$ with $\rho_{1}(t)=[ \rho_{1,1}(t),...,\rho_{1,m}(t)  ]^\top$ be the vector of user-designed performance functions, with $\rho_{1,j}(t)>0$ being \textcolor{black}{differentiable}, decreasing and bounded, satisfying $\rho_{1,j}(0)>|x_{1,j}(0)-y_{d,j}(0)|$. Moreover, their first-order  derivatives $\dot \rho_{1,j}(t)\in\mathbb R$ are \textcolor{black}{piecewise continuous}, bounded but not necessarily known.  A standard selection of these performance functions is 
\begin{flalign}
	\rho_{1,j}(t)=(\rho_{1,j,0}-\rho_{1,j,\infty})\exp^{-l_{1,j}t}+\rho_{1,j,\infty}
\end{flalign}
where $\rho_{1,j,0}$, $l_{1,j}$, $\rho_{1,j,\infty}>0$ are design parameters that determine the maximum overshoot, minimum convergence rate and maximum steady-state accuracy for $e_j(t)$, respectively, and $\rho_{1,j,0}$ satisfies $\rho_{1,j,0}>|x_{1,j}(0)-y_{d,j}(0)|$ and $\rho_{1,j,0}>\rho_{1,j,\infty}$.
Then, the goal to ensure performance constraints is mathematically formulated as
\begin{flalign}
	|e_{j}(t)|<\rho_{1,j}(t),\ \ \forall j\in\mathbb I_1^m,\ \  \forall t\ge 0. \label{pc}
\end{flalign}

%In the absence of input saturation  (\ref{sat}), the standard PPC methodology \cite{BECHLIOULIS20141217} can be employed to enforce the performance constraints (\ref{pc})  for system (\ref{sys}). However, 
When input saturation (\ref{sat}) is present,  PPC \cite{BECHLIOULIS20141217} fails to achieve the satisfaction of performance constraints (\ref{pc}), 
since the control effort demanded by PPC cannot be delivered under insufficient actuator capacity, causing the output error to reach the performance boundary in finite time and triggering controller instability (refer to, Section II-B of \cite{10273607}). %since the internal instability phenomenon could occur. We refer the readers to Section II-B of \cite{10273607} for more details about this point.
As noted in Section \ref{sec:introduction}, existing results %to address such issue 
are restrictive. Therefore, we construct a bidirectional modification mechanism to address such an issue. % that can not only relax  the constraints when input saturation (\ref{sat}) occurs, but also accelerate the recovery of original constraints after saturation ceases, and further tighten the constraints to enhance  performance if saturation remains inactive at the steady-state phase.
%\textcolor{red}{that relaxes the performance functions $\rho_1(t)$ as little as possible when saturation is severe, and accelerates the tightening of $\rho_1(t)$ when saturation is mild,  %total control  magnitude $\sum_{j=1}^{m}|u_j(t)|$ 
%and the $\mathcal L_1$ norm of the control input   $\|u(t)\|_1$ falls below a prescribed threshold, 
%rendering the control task feasible and the performance degradation alleviated. Moreover, the output tracking performance can be further enhanced if saturation is generally mild.}
%we will construct a dynamical modification mechanism to appropriately adjust the performance functions $\rho_1(t)$ whenever input saturation (\ref{sat}) becomes active, thereby rendering the control task feasible. Moreover,  this mechanism also adjusts $\rho_1(t)$ whenever the control norm $\|u(t)\|$  falls below a prescribed design parameter, to further enhance the output tracking performance.
Let  %$\sigma(t)=[  \sigma (t),...,\sigma (t) ]^\top\in \mathbb R^m$ 
$\sigma(t)\in \mathbb R$ be the  modification signal for performance functions $\rho_1(t)$. Given the aforementioned background, the control problem considered in this paper is stated as follows. 

\textit{Control Problem:} Consider %the high-order uncertain  highly-coupled MIMO  strict-feedback nonlinear 
system (\ref{sys})  subject to input saturation (\ref{sat}) and performance constraints. Construct a \textcolor{black}{novel constraint modification mechanism} to alleviate potential conflict while reducing conservatism of control performance in existing methods.
% so that input saturation (\ref{sat}) is mitigated while performance degradation is alleviated, and the tracking performance can be further enhanced if saturation is mostly inactive. %to simultaneously alleviate the adverse influence of input saturation (\ref{sat}) and enhance the tracking performance specifications when the control norm $\|u(t)\|$ is low. 
Based on it, design a continuous, low-complexity state-feedback controller, so that: 
%\begin{enumerate}
	%\item%[\textbf{(i)}]  
	%The following tracking performance constraints on the output tracking errors $e(t)$ are satisfied, i.e.,
	1) the output tracking errors $e(t)$ satisfy modified constraints
	\begin{flalign}
		|e_j(t)|<\textcolor{black}{\rho_{1,j}(t)+\frac{c_{1,j}}{\mu} \sigma(t)},\ \ \forall j\in\mathbb I_1^m,\ \ \forall t\ge 0 \label{PC}
	\end{flalign}
	where $c_{1,j},\mu>0$ are design parameters;  %associated with the first-order normalized errors $\xi_{1,j}(t)$ as defined in (\ref{xi1jt}), and the dynamics of  modification signal $\sigma(t)$ as designed in (\ref{dsigma}), respectively;
	%\item%[\textbf{(ii)}] 
	and 2) All  signals in the closed-loop  are bounded, $\forall t\ge 0$.
	%\item%[\textbf{(iii)}] The modification signal $\sigma(t)$'s dynamics is of first-order, thus limiting the overall complexity of the controller;
	%\item The control design involves neither prior knowledge of the saturation level $\bar u_j$ or any feasibility condition, nor any derivative of the desired tracking trajectories $y_d(t)$.
%\end{enumerate}
%To obtain the solution to the control problem, an assumption besides the Assumptions 1-2 is introduced.

\textcolor{black}{\textit{Assumption 3:}} The desired tracking trajectories $y_d(t)$ are differentiable and bounded, $\forall t\ge 0$. Moreover, their first-order   derivatives $\dot y_d(t)=[ \dot y_{d,1}(t),...,\dot y_{d,m}(t)  ]^\top\in\mathbb R^m$ are \textcolor{black}{piecewise continuous}, bounded but not necessarily known, %To facilitate the stability analysis, we assume that 
  %\textcolor{black}{$|y_{d,j}(t)|\le\bar y_{d,j}$} and 
satisfying \textcolor{black}{$|\dot y_{d,j}(t)|\le\bar {\dot y}_{d,j}$}, $\forall t\ge 0$ for  unknown constants $%\bar y_{d,j},
\bar {\dot y}_{d,j}\ge0$.

%\textit{Remark x:} xxx.

\section{Controller Design}
\label{sec:results}
\subsection{Bidirectional Constraint  Modification Mechanism}
The modification signal $\sigma(t)$ %, $\sigma(0)=0$ 
of the performance functions $\rho_1(t)$ is generated by a  \textcolor{black}{single first-order dynamics:}
\begin{flalign}
        %&\dot \sigma(t)= -\beta \sigma(t) + \mu |\Delta u(t)|+ \gamma \Delta \hat u(t),\ \ \sigma(0)=0_m\notag\\
    \dot \sigma(t)= -\beta \sigma(t) + %\mu\sum_{j=1}^{m} |\Delta u_j(t)|
    \mu\|\Delta u(t)\|_1+ \gamma \Delta \hat u(t),\ \ \sigma(0)=0
        \label{dsigma}
\end{flalign}
where %$\beta=\operatorname{diag}\left(\beta_{1},...,\beta_{m} \right)\in\mathbb R^{m\times m}$, $\mu=\operatorname{diag}\left( \mu ,...,\mu \right)\in\mathbb R^{m\times m}$ and $\gamma=\operatorname{diag}\left(\gamma_1,...,\gamma_m \right)\in\mathbb R^{m\times m}$ with 
$\beta ,\mu,\gamma > 0$ are design parameters,  $\Delta u(t)=[ \Delta u_1(t),...,\Delta u_m(t)  ]^\top\in\mathbb R^m$  with 
$\Delta u_j(t)\in\mathbb R$, $j\in\mathbb I_1^m$ being the saturation deficiency   defined as
\begin{flalign}
	\Delta u_{j}(t):=\operatorname{sat}(u_{j}(t))-u_{j}(t) \label{dujt}
\end{flalign}
%$|\Delta u(t)|:=[ |\Delta u_1(t)|,...,|\Delta u_m(t)|  ]^\top\in\mathbb R^m$, 
and %$\Delta \hat u(t)=[ \Delta \hat u_1(t),...,\Delta \hat u_m(t)  ]^\top\in\mathbb R^m$ with 
$\Delta \hat u(t)\in\mathbb R$ is the performance enhancement term with
\begin{flalign}
    \Delta \hat u(t):=\left\{ \begin{array}{r}
	 \|u(t)\|_1-\hat u,\ \   \text{if}\  \|u(t)\|_1< \hat u \\
    0,\ \ \text{if}\ \|u(t)\|_1\ge \hat u \\
	\end{array} \right. \label{huj}
\end{flalign}
\iffalse
\begin{flalign}
	%&\Delta \hat u (t):=\left\{ \begin{array}{r}
		%|u_{j}(t)|-\hat u ,\ \  \text{if}\ |u_{j}(t)|< \hat u \\
		%0,\ \  \text{if}\ |u_{j}(t)|\ge \hat u \\
	%\end{array} \right.\notag\\
	\Delta \hat u(t):=\left\{ \begin{array}{l}
		%\sum_{j=1}^{m}|u_j(t)|
		\|u(t)\|_1-\hat u,\ \   \text{if}\ %t\ge t_s \text{and} 
			%\|u(t)\|_1< \hat u
		|u_j(t)|<\hat u_j,\ \forall j\in\mathbb I_1^m	\\
		0,\ \  %\text{if}\ \|u(t)\|_1\ge \hat u
		\text{otherwise}
		\\
	\end{array} \right. 
\end{flalign}\fi 
where %$0<\hat u \le \bar u_{j}$
%\textcolor{black}{$0<\hat u_j\le \bar u_j$} are design parameters and \textcolor{black}{$\hat u:=\sum_{j=1}^{m}\hat u_j$.} 
$\hat u>0$ is a design parameter. Then, it is derived from (\ref{huj}) that
\begin{flalign}
	-\hat u \le\Delta\hat u (t)\le 0,\ \ \forall t\ge 0. \label{dhu}
\end{flalign}

\textcolor{black}{\textit{Remark 2:}} When input saturation (\ref{sat}) occurs, the deficiency term $\|\Delta u(t)\|_1$ will dominate the dynamics (\ref{dsigma}), causing $\sigma(t)$ to increase and thus relaxing the performance constraints to alleviate potential conflict. Conversely, after saturation ceases,
the non-positive term $\Delta \hat u(t)$ will accelerate the decrease of $\sigma(t)$, thereby speeding up the recovery of original constraints.
Moreover, if saturation remains inactive at the steady-state phase (i.e., the interval $[t_{j}^*,+\infty)$ where $t_{j}^*:=\inf\{t\ge 0| \rho_{1,j}(t)\le 1.02\rho_{1,j,\infty}\}$ is the time when $\rho_{1,j}(t)$ first enters the $2\%$ band of its steady-state value), %\footnote{Throughout the article, we refer to the \textit{steady-state phase} as the interval $[t_{j}^*,+\infty)$ where $t_{j}^*:=\inf\{t\ge 0| \rho_{1,j}(t)\le 1.02\rho_{1,j,\infty}\}$ is the time when the performance function first enters the $2\%$ band of its steady-state value.}, 
$\Delta \hat u(t)$ will make the constraints further tightened, thus enhancing control performance. %without incurring any cost.

\textit{Remark 3:} In \cite{ji2021saturation,ji2023saturation,11304721,10273607}, the order of modification dynamics must match that of the  plant. Although first-order modification schemes are developed in \cite{11130930,trakas2024adaptive}, a separate dynamics is required for each dimensional constraint, and the plants are restricted to MIMO normal-form  systems.
In contrast, the dynamics (\ref{dsigma}) is both first-order and single, %whose output simultaneously adjusts all constraints, 
thereby significantly reducing the complexity  introduced into controller and increasing the flexibility of input-saturated PPC design. Moreover, the dynamics (\ref{dsigma}) can also be designed using the $\mathcal{L}_2$ norm; however, it would increase the computational burden due to the involved square-root and  multiplication operations.

\iffalse
\textcolor{black}{Let  $\sigma=[\sigma,...,\sigma_m]^\top$ be generated by}
\begin{flalign}
	\dot \sigma=-\beta \sigma+\operatorname{sgn}(\hat e)\operatorname{sgn}(\hat g)\Delta u+\operatorname{sgn}(\hat e)\Delta \bar u,\ \ \
	\sigma(0)=0_m \label{dsigma}
\end{flalign}
where $\beta=\operatorname{diag}(\beta_1,...,\beta_m)$ with $\beta_i>0$, $\operatorname{sgn}(\hat e)=\operatorname{diag}(\operatorname{sgn}(e_{1}),...,\operatorname{sgn}(e_{m}))$, $\operatorname{sgn}(\hat g)=\operatorname{diag}(\operatorname{sgn}(g_{1,1}),...,\operatorname{sgn}(g_{m,m}))$, $\Delta u=[\Delta u_1,...,\Delta u_m]^\top=\operatorname{sat}(u)-u$ with $\Delta u_i=\operatorname{sat}(u_i)-u_i$, $\Delta \bar u=[\Delta \bar u_1,...,\Delta \bar u_m]^\top=\operatorname{sgn}(u)\operatorname{sat}(u)-\bar u$, $\operatorname{sgn}(u)=[\operatorname{sgn}(u_1),...,\operatorname{sgn}(u_m)]^\top$, $\bar u=[\bar u_1,...,\bar u_m]^\top$.\fi

\subsection{Low-Complexity Control Scheme Design}
\label{sCD}
%\textcolor{red}{\textit{Assumption 4.1:}} The diagonal entries $g_{n,j,j}$ of $g_{n}$ are either strictly positive or strictly negative, and their  signs are known. Moreover, there exist unknown constants \textcolor{black}{$\underline g_{n,j,j}>0$} so that $|g_{n,j,j}|\ge\underline g_{n,j,j}$.
 
%In standard PPC design, the performance functions $\rho_{1,j}(t)$ are used to normalize the output tracking errors $e_j(t)$, i.e., $\xi_{1,j}(t)=\frac{e_j(t)}{\rho_{1,j}(t)}$, where $\xi_{1,j}(t)$ are the normalized errors. $\xi_{i,j}(t)$. 
By following the design philosophy of PPC %methodology 
\cite{BECHLIOULIS20141217}, a backstepping-like control design program is adopted in this paper. Throughout such design, a \textcolor{black}{differentiable} nonlinear mapping function $T:(-1,1)\rightarrow \mathbb R$ is leveraged, satisfying the following \textcolor{black}{properties:} 1) $T(0)=0$,  $\lim_{\diamond\rightarrow \pm1}T(\diamond)=\pm\infty$; and 2) $[dT(\diamond)/d\diamond]$ is \textcolor{black}{locally Lipschitz} in $\diamond$, with $0<[dT(\diamond)/d\diamond]<+\infty$ and $\lim_{\diamond\rightarrow \pm1} [dT(\diamond)/d\diamond]=+\infty$.
\iffalse
	\begin{enumerate}
		\item%[\textbf{(i)}]
		 $T(0)=0$, $\lim_{\diamond\rightarrow \pm1}T(\diamond)=\pm\infty$; %, and $\operatorname{sgn}(T(\diamond))=\operatorname{sgn}(\diamond)$;
		\item%[\textbf{(ii)}]
		 $[dT(\diamond)/d\diamond]$ is \textcolor{black}{locally Lipschitz} in $\diamond$, with $0<[dT(\diamond)/d\diamond]<+\infty$ and $\lim_{\diamond\rightarrow \pm1} [dT(\diamond)/d\diamond]=+\infty$. %, and \textcolor{black}{$\lim_{\diamond\rightarrow \pm1}[d^2T(\diamond)/d\diamond^2]=\pm\infty$.} 
	\end{enumerate}\fi 
%Denote the \textcolor{black}{inverse function} of $T(\diamond)$ by $T^{-1}:\mathbb R\rightarrow (-1,1)$. 
A feasible selection of it %the nonlinear mapping function 
is $T(\diamond)=\ln\left(\frac{1+\diamond}{1-\diamond} \right)$.
\iffalse
\begin{flalign}
	T(\diamond)=\ln\left(\frac{1+\diamond}{1-\diamond} \right),\ \ T^{-1}(\diamond)=\frac{\exp^\diamond-1}{\exp^\diamond+1}.
	%T(\diamond)=\tan\left(\frac{\pi}{2}\diamond \right),\ \ T^{-1}(\diamond)=\frac{2}{\pi}\arctan\left(\diamond \right).
\end{flalign}\fi
With the introduction of  %function  
$T(\diamond)$, the low-complexity controller is designed as follows.

\textbf{Step $1$:} Let $\xi_1(t)=[ \xi_{1,1}(t),...,\xi_{1,m}(t)  ]^\top\in\mathbb R^m$ be the vector of first-order normalized errors, with $\xi_{1,j}(t)$, $j\in\mathbb I_1^m$ defined as
\begin{flalign}
	\xi_{1,j}(t):=\frac{x_{1,j}(t)-\alpha_{0,j}(t)}{\textcolor{black}{\rho_{1,j}(t)+\eta_{1,j}  \sigma(t)}} \label{xi1jt}
\end{flalign}
where $\alpha_{0,j}(t):=y_{d,j}(t)$, $\eta_{1,j}:=\frac{c_{1,j}}{\mu}$ and $c_{1,j}>0$ are design parameters. Let $\epsilon_1(t)=[ \epsilon_{1,1}(t),..,\epsilon_{1,m}(t)  ]^\top\in\mathbb R^m$  be the vector of first-order  mapped errors, with $\epsilon_{1,j}(t)$ defined as
\begin{flalign}
	\epsilon_{1,j}(t):=T(\xi_{1,j}(t)).\label{ep1j}
\end{flalign}
%,  and $\Xi_1=\operatorname{diag}\left( \Xi_{1,1},...,\Xi_{1,m}\right)$ with $\Xi_{1,j}:=\frac{d T(\xi_{1,j})/d \xi_{1,j}}{\rho_{1,j}+\eta_{1,j} \sigma}$.
Design the first-order  virtual controller $\alpha_1(t)=[ \alpha_{1,1}(t),...,\alpha_{1,m}(t)  ]^\top\in\mathbb R^m$ as
\begin{flalign}
	\alpha_1(t):=-k_1%\Xi_1
	\epsilon_1(t) \label{a1}
\end{flalign}
where $\textcolor{black}{k_1}>0$ is a design parameter. %\textcolor{black}{$k_1=\operatorname{diag}\left(k_{1,1},k_{1,2}  \right)$}

\textbf{Step $i\in\mathbb{I}_2^{n}$:} Let $\xi_i(t)=[  \xi_{i,1}(t),...,\xi_{i,m}(t) ]^\top\in\mathbb R^m$ be the vector of $i$th-order normalized errors with $\xi_{i,j}(t)$, $j\in\mathbb I_1^m$ defined as
\begin{flalign}
	\xi_{i,j}(t):=\frac{x_{i,j}(t)-\alpha_{i-1,j}(t)}{\rho_{i,j}(t)+\eta_{i,j}  \sigma(t)} \label{xiijt}
\end{flalign}
where $\eta_{i,j}:=\frac{c_{i,j}}{\mu}$, $c_{i,j}>0$ are design parameters, and  $\rho_{i,j}:\mathbb{R}_{\ge 0}\to\mathbb{R}$ are any strictly positive, \textcolor{black}{ differentiable} and bounded functions, satisfying $\rho_{i,j}(0)>|x_{i,j}(0)-\alpha_{i-1,j}(0)|$. \textcolor{black}{Moreover, their first-order  derivatives $\dot\rho_{i,j}(t)\in\mathbb R$ are \textcolor{black}{piecewise continuous} and bounded.} %but not necessarily known. 
Typically, these functions are selected %in the same form of 
similarly to $\rho_{1,j}(t)$: 
\begin{flalign}
	\rho_{i,j}(t)=(\rho_{i,j,0}-\rho_{i,j,\infty})\exp^{-l_{i,j}t}+\rho_{i,j,\infty}
\end{flalign}
where $\rho_{i,j,0},l_{i,j},\rho_{i,j,\infty}>0$ are design parameters, and $\rho_{i,j,0}$ satisfies $\rho_{i,j,0}>|x_{i,j}(0)-\alpha_{i-1,j}(0)|$ and $\rho_{i,j,0}>\rho_{i,j,\infty}$.
Then, let $\epsilon_i(t)=[  \epsilon_{i,1}(t),..,\epsilon_{i,m}(t) ]^\top\in\mathbb R^m$  be the vector of $i$th-order  mapped errors, with $\epsilon_{i,j}(t)$  defined as
\begin{flalign}
	\epsilon_{i,j}(t):=T(\xi_{i,j}(t)). \label{epij}
\end{flalign}
%, and $\Xi_i=\operatorname{diag}\left(\Xi_{i,1},...,\Xi_{i,m} \right)$ with $\Xi_{i,j}:=\frac{d T(\xi_{i,j})/d \xi_{i,j}}{\rho_{i,j}+\eta_{i,j} \sigma}$.
Design the $i$th-order  virtual
controller $\alpha_i(t)=[ \alpha_{i,1}(t),...,\alpha_{i,m}(t)  ]^\top\in\mathbb R^m$ as
\begin{flalign}
	\alpha_i(t)=-k_i%\Xi_i
	\epsilon_i(t) \label{ai}
\end{flalign}
where $k_i>0$ are design parameters \textcolor{black}{and $\alpha_n(t)=u(t)$.}

\iffalse
\textbf{Step $n+1$:} %Let $\xi_n=[ \xi_{n,1},...,\xi_{n,m}  ]^\top$ with $\xi_{n,j}:=\frac{x_{n,j}-\alpha_{n-1,j}}{\rho_{n,j}+\eta_{n,j} \sigma}$, and $\epsilon_n=[  \epsilon_{n,1},..,\epsilon_{n,m} ]^\top$  with  $\epsilon_{n,j}:=T(\xi_{n,j})$. %, and $\Xi_n=\operatorname{diag}\left(\Xi_{n,1},...,\Xi_{n,m} \right)$ with $\Xi_{n,j}:=\frac{d T(\xi_{n,j})/d \xi_{n,j}}{\rho_{n,j}+\eta_{n,j} \sigma}$.
Now, design the %$n$th order virtual controller $\alpha_n$ and the  unsaturated 
actual controller $u(t)$ as
\begin{flalign}
	u(t)=\alpha_n(t)=-k_n%\Xi_n
	\epsilon_n(t). \label{u}
\end{flalign}\fi
%where  $k_{n}>0$ is a design parameter. 
\iffalse Denote the modified performance functions by $\hat \rho_{i}(t)=[  \hat \rho_{i,1}(t),...,\hat \rho_{i,m}(t) ]^\top$ with
\begin{flalign}
	\hat \rho_{i,j}(t):=\rho_{i,j}(t)+\eta_{i,j} \sigma(t)
\end{flalign}
where  $\rho_{i}(t)=[ \rho_{i,1}(t),...,\rho_{i,m}(t)  ]^\top$ are designed performance functions with $\rho_{i,j}(t)>0$ being \textcolor{red}{continuously differentiable, bounded, and decreasing functions of time.} One typical selection of these performance functions is $\rho_{i,j}(t)=(\rho_{i,j,0}-\rho_{i,j,\infty})\exp^{-l_{i,j}t}+\rho_{i,j,\infty}$, \textcolor{black}{where $\rho_{i,j,0}>|x_{i,j}(0)-y_{d,j}(0)|$} and $l_{i,j},\rho_{i,j,\infty}>0$ are design parameters.\fi

In each step of the backstepping-like design program, % (excluding Step $n+1$), 
owing to the construction $\rho_{i,j}(0)>|x_{i,j}(0)-\alpha_{i-1,j}(0)|$, $\forall i\in\mathbb{I}_1^{n}$, $\forall j\in\mathbb{I}_1^{m}$ and $\sigma (0)=0$ in (\ref{dsigma}), we know that $|\xi_{i,j}(0)|<1$, $\forall i\in\mathbb{I}_1^{n}$, $\forall j\in\mathbb{I}_1^{m}$. Therefore, $\epsilon_i(0)$, $\forall i\in\mathbb{I}_1^{n}$ are well-defined. On the other hand, since $\rho_{i,j}(0)+\eta_{i,j}  \sigma(0)=\rho_{i,j}(0)>0$, we need to guarantee that
\begin{flalign}
	\rho_{i,j}(t)+\eta_{i,j}  \sigma(t)>0,\ \  \forall i\in\mathbb{I}_1^{n},\ \ \forall j\in\mathbb{I}_1^{m},\ \  \forall t\ge 0 \label{rijt}
\end{flalign} 
to avoid the occurrence of $\rho_{i,j}(t)+\eta_{i,j} \sigma(t)=0$.
Let $\Delta(t) :=\mu\|\Delta u(t)\|_1+\gamma \Delta \hat u (t)$. Then, it follows from (\ref{dsigma}), the facts $\|\Delta u(t)\|_1\ge 0$, $\forall t\ge 0$ and $-\hat u \le\Delta\hat u (t)\le 0$, $\forall t\ge 0$ that
\begin{flalign}
	\sigma(t)=&\ \exp^{-\beta t}\sigma(0)+\exp^{-\beta t}\int_{0}^{t}\exp^{\beta \tau}\Delta (\tau)d\tau \notag\\
	\ge&\ \exp^{-\beta t}  \int_{0}^{t} \exp^{\beta \tau}\gamma \Delta \hat u (\tau)d\tau \ge -\exp^{-\beta t}  \int_{0}^{t} \exp^{\beta \tau}\gamma  \hat u  d\tau
	\notag\\
	=&\ -\frac{\gamma  \hat u }{\beta }(1-\exp^{-\beta t}) \ge -\frac{\gamma  \hat u }{\beta }:=\underline\sigma,\ \  \forall t\ge 0. \label{usj}
\end{flalign}
Since $\rho_{i,j}(t)+\eta_{i,j} \sigma (t)\ge \rho_{i,j,\infty}+\eta_{i,j}  \underline\sigma$, $\forall i\in\mathbb{I}_1^n$, $\forall j\in\mathbb{I}_1^m$, $\forall t\ge 0$, to make (\ref{rijt}) hold, the design parameters should be chosen such that $\rho_{i,j,\infty}+\eta_{i,j}  \underline\sigma>0$, that is,
\begin{flalign}
	c_{i,j}\gamma  \hat u <\mu\beta \rho_{i,j,\infty}, \ \ \forall i\in\mathbb{I}_1^n,\ \ \forall j\in\mathbb{I}_1^m. \label{cij}
\end{flalign}

\textcolor{black}{\textit{Remark 4:}} As can be seen from (\ref{xi1jt})-(\ref{ai}), relative to the standard PPC framework \cite{BECHLIOULIS20141217}, our design introduces only a scalar modification signal $\sigma(t)$ governed by the first‑order dynamics (\ref{dsigma}). Consequently, the additional complexity imposed on the controller is significantly reduced and the flexibility of input-saturated PPC design is increased, compared with existing approaches that either rely on higher‑order modification dynamics \cite{ji2021saturation,ji2023saturation,11304721,10273607} or involve multiple first‑order generation dynamics for MIMO systems \cite{11130930,trakas2024adaptive}. Moreover, the design parameters $\mu$ and $c_{i,j}$ are purposefully embedded into the controller design, which not only renders the satisfaction of modified constraints (\ref{PC}) feasible, but also, as will be shown in the  stability analysis, enables an explicit trade-off between modification magnitude and allowable tightening of performance constraints. %enabling a balance between the amplitude of modification signal $\sigma(t)$ and the maximum attainable performance enhancement.
%It can be seen from (\ref{xi1jt}) and (\ref{xiijt}) that, on the basis of standard PPC \cite{BECHLIOULIS20141217}, our design only introduces the scalar modification signal $\sigma(t)$, which is of first-order and thus minimizes the additional complexity introduced into the controller, as compared with the results that involve higher-order modification dynamics \cite{ji2021saturation,ji2023saturation,11304721,10273607} or multiple first-order generation dynamics \cite{11130930,trakas2024adaptive} for MIMO systems. Moreover, the design parameters $\mu$ and $c_{i,j}$ are deliberately introduced into the design, which make the satisfaction of modified performance constraints feasible and  the magnitude of modification signal against the maximum achievable performance enhancement can be balanced, as seen in the subsequent stability analysis.

%\textit{Remark x:} $k_i,\textcolor{black}{K_n}$, and $\Xi_i$.

\iffalse
\textit{Remark x:} If $\gamma $ or $\hat u $ is set as zero, then (\ref{cij}) holds true naturally. At the same time, the term $\gamma \Delta\hat u (t)$ in (\ref{dsigma})  will not take effect. Therefore, (\ref{dsigma}) reduces to
\begin{flalign}
		\dot \sigma(t)= -\beta \sigma(t) + \mu |\Delta u(t)|,\ \ \sigma(0)=0_m.\label{dst}
\end{flalign}
Solving (\ref{dst}) results in
\begin{flalign}
	\sigma(t)&= \exp^{-\beta t}\int_{0}^{t}\exp^{\beta \tau}\mu |\Delta u_j(\tau)| d\tau\ge 0\notag\\
	&\ \ \ \forall j\in\mathbb{I}_1^m,\ \ \forall t\ge 0.
\end{flalign}
Therefore, (\ref{rijt}) holds true naturally. This implies that xxx.
\fi 

\section{Stability Analysis}
Our main results are concluded into the following theorem.

\textcolor{black}{\textit{Theorem 1:}} Consider %the high-order uncertain highly-coupled MIMO  strict-feedback nonlinear 
system (\ref{sys}) subject to input saturation  (\ref{sat}) and %output tracking 
performance constraints (\ref{PC}). Let Assumptions 1-3 hold. Choose the %design 
parameters $c_{i,j},\gamma, \hat u, \mu,\beta ,\rho_{i,j,\infty}>0$ such that the inequality
\begin{flalign}
	\textcolor{black}{c_{i,j}\gamma  \hat u <\mu\beta \rho_{i,j,\infty},\ \ \forall i\in\mathbb{I}_1^{n},\ \ \forall j\in\mathbb{I}_1^m} \label{c1}
\end{flalign}
is satisfied.  If the control scheme %developed  
in Section \ref{sec:results} is applied, then there exist unknown  constants $K_{n-1,j}>0$, independent of $c_{n,j}$, such that the control problem %stated 
in  Section \ref{PF} is solved for all $c_{n,j}$ satisfying
\begin{flalign}
	\textcolor{black}{c_{n,j}>K_{n-1,j},\ \ \forall j\in\mathbb{I}_1^m.} \label{c2}
\end{flalign}

\textit{Proof:} First, we transform the dynamics of $x_i(t)$, $i\in\mathbb{I}_1^n$, i.e., system (\ref{sys}), into the dynamics of $\xi_i(t)$. Let $\xi_i^\ast(t)=\operatorname{diag}\left(\xi_{i,1}(t),...,\xi_{i,m}(t)\right)\in\mathbb R^{m\times m}$ and $\eta_i= [ \eta_{i,1},...,\eta_{i,m}  ]^\top \in\mathbb R^{m}$. % with $\eta_{i,j}>0$, $j\in\mathbb{I}_1^m$ being defined as
From the definition of $\xi_i(t)$ in %Section \ref{sCD}
(\ref{xi1jt}) and (\ref{xiijt}), we have
\begin{flalign}
	x_1(t)&=\xi_1^*(t)\left(\rho_1(t)+\eta_1\sigma(t) \right)+y_d(t)\notag\\ x_i(t)&=\xi_i^*(t)\left(\rho_i(t)+\eta_i\sigma(t) \right)+\alpha_{i-1}(t),\ \ i\in\mathbb{I}_2^n \label{xi}
\end{flalign}
where $\rho_i(t):=[ \rho_{i,1}(t),...,\rho_{i,m}(t)  ]^\top\in\mathbb{R}^m$.
Let $\Theta_i(t)=\operatorname{diag}\left(\frac{1}{\rho_{i,1}(t)+\eta_{i,1} \sigma (t)},...,\frac{1}{\rho_{i,m}(t)+\eta_{i,m} \sigma (t)}\right)\in\mathbb R^{m\times m}$, $i\in\mathbb{I}_1^n$, let $\tilde\xi_i(t)=[ \xi_1^\top(t),...,\xi_i^\top(t) ]^\top\in\mathbb R^{mi}$, and let $\dot\rho_i(t)=[ \dot\rho_{i,1}(t),...,\dot\rho_{i,m}(t)  ]^\top\in\mathbb R^{m}$. From (\ref{ep1j}), (\ref{a1}), (\ref{epij}) and (\ref{ai}), $\dot\alpha_{i,j}(t)$, $i\in\mathbb{I}_1^{n-1}$,  $j\in\mathbb{I}_1^{m}$ are calculated as
\begin{flalign}
	\dot\alpha_{i,j}(t)=-k_i\frac{dT(\xi_{i,j}(t))}{d\xi_{i,j}(t)}\dot\xi_{i,j}(t).\label{dai}
\end{flalign}
By using (\ref{xi1jt}), (\ref{xiijt}), (\ref{sys}), (\ref{xi})-(\ref{dai}), the dynamics of $\xi_i(t)$, $i\in\mathbb{I}_1^{n}$ is:
\begin{flalign}
	& \dot \xi_1(t):=  h_1(\tilde\xi_2(t),\xi_n(t),\sigma(t),t) =  \Theta_1(\sigma(t),t)\Big(f_1(\xi_1(t),\sigma(t),t)\notag\\&
	+g_1(\xi_1(t),\sigma(t),t) \big(\alpha_1(\xi_1(t))
	+\xi_2^*(t)\rho_2(t)
	+\xi_2^*(t)\eta_2\sigma(t)\big)\notag\\&-\dot y_d(t)
	-\xi_1^*(t)(\dot\rho_1(t)+\eta_1\textcolor{black}{\dot\sigma(\xi_n(t),\sigma(t))})\Big)\notag\\
	& \dot \xi_i(t):=  h_i(\tilde\xi_{i+1}(t),\xi_n(t),\sigma(t),t) =  \Theta_i(\sigma(t),t)\Big(f_i(\tilde\xi_i(t),\sigma(t),t)\notag\\&+g_i(\tilde\xi_i(t),\sigma(t),t) \big(\alpha_i(\xi_i(t))
	+\xi_{i+1}^*(t)(\rho_{i+1}(t)+\eta_{i+1}\sigma(t))
	\big)\notag\\&
	-\textcolor{black}{\dot \alpha_{i-1}(\tilde\xi_{i}(t),\xi_n(t),\sigma(t),t)}-\xi_i^\ast(t)(\dot\rho_i(t) 
	+\eta_i\dot\sigma(\xi_n(t),\sigma(t)))\Big)%, i\in\mathbb{I}_2^{n-1}
	\notag\\ & \dot \xi_n(t):=  h_n(\tilde\xi_{n}(t),\sigma(t),t) =  \Theta_n(\sigma(t),t)\Big(f_n(\tilde\xi_n(t),\sigma(t),t)\notag\\&+g_n(\tilde\xi_n(t),\sigma(t),t) \operatorname{sat}\big(u(\xi_n(t))\big)
	-\textcolor{black}{\dot \alpha_{n-1}(\tilde\xi_n(t),\sigma(t),t)}\notag\\&-\xi_n^\ast(t)(\dot\rho_n(t)+\eta_n\dot\sigma(\xi_n(t),\sigma(t)))\Big).\label{dxii}
\end{flalign}
\iffalse
\begin{flalign}
	\dot \xi_1(t):=&\ h_1(\tilde\xi_2(t),\xi_n(t),\sigma(t),t)\notag\\
	=&\ \Theta_1(\sigma(t),t)\Big(f_1(\xi_1(t),\sigma(t),t)+g_1(\xi_1(t),\sigma(t),t)\notag\\&\times\big(\alpha_1(\xi_1(t))
	+\xi_2^*(t)\rho_2(t)
	+\xi_2^*(t)\eta_2\sigma(t)\big)-\dot y_d(t)\notag\\&
	-\xi_1^*(t)(\dot\rho_1(t)+\eta_1\textcolor{black}{\dot\sigma(\xi_n(t),\sigma(t))})\Big)\notag\\
	\dot \xi_i(t):=&\ h_i(\tilde\xi_{i+1}(t),\xi_n(t),\sigma(t),t) \notag\\
	=&\ \Theta_i(\sigma(t),t)\Big(f_i(\tilde\xi_i(t),\sigma(t),t)+g_i(\tilde\xi_i(t),\sigma(t),t)\notag\\&
	\times\big(\alpha_i(\xi_i(t))
	+\xi_{i+1}^*(t)\rho_{i+1}(t)
	+\xi_{i+1}^*(t)\eta_{i+1}\sigma(t)\big)\notag\\&
	-\textcolor{black}{\dot \alpha_{i-1}(\tilde\xi_{i}(t),\xi_n(t),\sigma(t),t)}-\xi_i^\ast(t)(\dot\rho_i(t)\notag\\&
	+\eta_i\dot\sigma(\xi_n(t),\sigma(t)))\Big),\ \ i\in\mathbb{I}_2^{n-1} \notag\\
	\dot \xi_n(t):=&\ h_n(\tilde\xi_{n}(t),\sigma(t),t)\notag\\
	=&\ \Theta_n(\sigma(t),t)\Big(f_n(\tilde\xi_n(t),\sigma(t),t)+g_n(\tilde\xi_n(t),\sigma(t),t)\notag\\&
	\times\operatorname{sat}\big(u(\xi_n(t))\big)
	-\textcolor{black}{\dot \alpha_{n-1}(\tilde\xi_n(t),\sigma(t),t)}\notag\\&-\xi_n^\ast(t)(\dot\rho_n(t)+\eta_n\dot\sigma(\xi_n(t),\sigma(t)))\Big).\label{dxii}
\end{flalign}
\fi 
Let $z(t)=[ {\tilde\xi}_n^\top(t),\sigma(t)  ]^\top\in\mathbb R^{mn+1}$, whose %. Then, the $z(t)$-
dynamics is
\begin{flalign}
	\dot z(t):=&\ F(z(t),t) 
	=
	\begin{bmatrix}
		h_1^\top\left(\tilde\xi_2(t),\xi_n(t),\sigma(t),t \right)\\
		\vdots\\
		h_n^\top\left( \tilde\xi_{n}(t),\sigma(t),t\right)\\
		\dot\sigma^\top\left( \xi_n(t),\sigma(t)\right)\\
	\end{bmatrix}. \label{dz} %[ h_1^\top(\tilde\xi_2(t),\xi_n(t),\sigma(t),t),...,h_n^\top(\tilde\xi_{n}(t),\sigma(t),t),\dot\sigma^\top(\xi_n(t),\sigma(t),t) ]^\top.\notag
\end{flalign}
For the $z(t)$-dynamics (\ref{dz}), we define a non-empty and open set $\Omega_z=(-1,1)^{mn}\times\mathbb R$. In the sequel, we proceed the proof in 3 stages.   In the first stage, we will show that there exists a unique and maximal solution $z:[0,t_{\max})\to\Omega_z$ for (\ref{dz}), where $0<t_{\max}\le+\infty$ is an unknown time instant. In the second stage, we will show that, $\forall t\in[0,t_{\max})$, the unique and maximal solution $z(t)$ for (\ref{dz}) remains strictly within a compact subset of $\Omega_z$, and all the closed-loop signals in (\ref{dz}) are \textcolor{black}{bounded}. In the third stage, we will show that there exists a contradiction if $t_{\max}<+\infty$, which thus extends the unique and maximal solution of (\ref{dz}) to $z:[0,+\infty)\to\Omega_z$. 
%\textit{Lemma 1: } If $\rho_{i,j}+\sigma >0$, then .

\textbf{Stage I:} %\textit{existence  and uniqueness of the maximal solution  $z:[0,t_{\max})\to\Omega_z$ for (\ref{dz}).} 
The set $\Omega_z$ is non-empty and open.  We have shown in the reasoning from (\ref{ai}) to (\ref{rijt}) that  $|\xi_{i,j}(0)|<1$, $\forall i\in\mathbb{I}_1^{n}$, $\forall j\in\mathbb{I}_1^{m}$. From (\ref{dsigma}), we know that $\sigma (0)=0$. Therefore, it holds that $z(0)\in\Omega_z$. Moreover, for all $i\in\mathbb I_1^n$ and $j\in\mathbb I_1^m$, the following results hold: 1) $\rho_{i}(t)$ are differentiable in $t$  and $\dot\rho_{i}(t)$ are piecewise continuous in $t$; 2)  $f_{i}(\tilde x_i,t)$ and $g_{i}(\tilde x_i,t)$ are locally Lipschitz in $\tilde x_i$ and piecewise continuous in $t$; 3) $\alpha_i(\xi_i(t))$  are differentiable in $\xi_i(t)$ since $T(\diamond)$ is  differentiable in $\diamond$; 4) $\operatorname{sat}\big (u(\xi_n(t)) \big )$ is Lipschitz in $u(\xi_n(t))$ since $\operatorname{sat}(\diamond)$ is Lipschitz in $\diamond$; 5) $\dot y_d(t)$ is piecewise continuous in $t$; 6)  ${dT(\xi_{i,j}(t))}/{d\xi_{i,j}(t)}$ are locally Lipschitz in $\xi_{i,j}(t)$ since ${dT(\diamond)}/{d \diamond}$ is  locally Lipschitz in $\diamond$; 7) $\|\Delta u(\xi_{n}(t))\|_1$ is Lipschitz in $\Delta u(\xi_{n}(t))$ \textcolor{black}{since $\|\diamond\|_1$ is Lipschitz in $\diamond$;} and 8)  $\Delta \hat u(\xi_n(t))$ is Lipschitz in $\|u(\xi_n(t))\|_1$ and $\|u(\xi_n(t))\|_1$ is Lipschitz in $u(\xi_n(t))$.

Since the addition, subtraction, multiplication, division (when well-defined) and composition preserve the properties of locally Lipschitz and piecewise continuous, with the results 1)-8), we can conclude that $F(z(t),t)$ is locally Lipschitz in $z(t)$ and piecewise continuous in $t$. Then, according to \textcolor{black}{Theorem 54 in \cite{sontag2013mathematical}}, the unique and maximal solution $z:[0,t_{\max})\to\Omega_z$ for (\ref{dz}) is ensured. Hence, it holds that
\begin{flalign}
	&|\xi_{i,j}(t)|<1,\ \ \forall i\in\mathbb I_1^n,\ \ \forall  j\in\mathbb I_1^m,\ \ \forall t\in[0,t_{\operatorname{max}})\notag\\
	&|\sigma (t)|<+\infty,\ \ \forall j\in\mathbb I_1^m,\ \ \forall t\in[0,t_{\operatorname{max}}). \label{uni}
\end{flalign}

\textbf{Stage II:} %\textit{strictly remaining of the unique and maximal solution $z(t)$ for (\ref{dz}) within a compact subset of $\Omega_z$, and the \textcolor{black}{boundedness} of all closed-loop signals in (\ref{dz}), $\forall t\in[0,t_{\max})$.} %If $t_{\max}<+\infty$, then there must exist some $i\in\mathbb{I}_1^n$ and $j\in\mathbb{I}_1^m$, such that $\lim_{t\to t_{\max}^-}|\xi_{i,j}(t)|=1$ and $\lim_{t\to t_{\max}^-}|\sigma(t)|=+\infty$.
To prove the former, we will show that, for all $i\in\mathbb I_1^{n}$ and $j\in\mathbb I_1^{m}$, there exist unknown constants $\bar\xi_{i,j}>0$ and $\bar\sigma>0$, such that $|\xi_{i,j}(t)|\le\bar\xi_{i,j}<1$ and $|\sigma(t)|\le\bar\sigma<+\infty$, $\forall t\in[0,t_{\max})$. From (\ref{uni}) and the continuity of $\xi_{i,j}(t)$ and $\sigma(t)$, it can be inferred that the proof equals to exclude the occurrence of $\lim_{t\to t_{\max}^-}|\xi_{i,j}(t)|=1$ and $\lim_{t\to t_{\max}^-}|\sigma(t)|=+\infty$,  $\forall i\in\mathbb I_1^{n}$ and $\forall j\in\mathbb I_1^{m}$. %To facilitate the readability of the proof, it  will be presented by the following 3 lemmas. %We first prove that the case $t_{\max}<+\infty$ cannot be attributed to the occurrence of $\lim_{t\to t_{\max}^-}|\sigma(t)|=+\infty$,

First, we exclude the occurrence of $\lim_{t\to t_{\max}^-}|\sigma(t)|=+\infty$, $\forall j\in\mathbb I_1^{m}$. %, as shown in \textcolor{black}{Lemma 3}.
%\textcolor{black}{\textit{Lemma 3:}} There must exist an unknown constant $\bar\sigma >0$ such that $|\sigma(t)|\le \bar\sigma <+\infty$, $\forall t\in[0,t_{\max})$.
%\textit{Proof:} 
We use the method of proof by contradiction. Owing to (\ref{usj}), we have $\sigma(t)\ge \underline\sigma $, $\forall t\ge 0$. This indicates that the occurrence of $\lim_{t\to t_{\max}^-}\sigma(t)=-\infty$ is impossible. Then, we only need to exclude the occurrence of $\lim_{t\to t_{\max}^-}\sigma(t)=+\infty$. To this end, we assume %there exist some $j\in\mathbb{I}_j\subset\mathbb{I}_1^m$ (where $\mathbb{I}_j\ne \{0\}$) such that
\begin{flalign}
	\lim_{t\to t_{\max}^-}\sigma(t)=+\infty. \label{hs}
\end{flalign} %Then, there must exist a sequence $\{t_{j}^k\}\subset [0,t_{\max})$, \textcolor{black}{$k\in\mathbb N\cup\{0\}$} such that $\sigma(t_j^k)\to +\infty$. 
From the BIBS stability of system (\ref{sys}) and the boundedness of $\operatorname{sat}(u(t))$ from (\ref{sat}), %i.e., $|\operatorname{sat}(u_j(t))|\le\bar u_j$, $\forall j\in\mathbb{I}_1^m$ and $\forall t\ge 0$, 
we know that, for all $i\in\mathbb{I}_1^n$ and $j\in\mathbb{I}_1^m$, $x_{i,j}(t)$ are bounded, $\forall t\ge 0$. In addition, from \textcolor{black}{Assumption 3} and construction, $y_{d,j}(t)$,  $j\in\mathbb{I}_1^m$ and $\rho_{i,j}(t)$, $i\in\mathbb{I}_1^n$, $j\in\mathbb{I}_1^m$ are bounded,  $\forall t\ge 0$. %Therefore, $e_{1,j}=x_{1,j}-y_{d,j}$ are bounded.
Therefore, it holds from (\ref{xi1jt}) that
\begin{flalign}
	\xi_{1,j}(t_{\max}^-)=\frac{x_{1,j}(t_{\max}^-)-y_{d,j}(t_{\max}^-)}{\rho_{1,j}(t_{\max}^-)+\eta_{1,j} \sigma(t_{\max}^-)}\to 0,\ \  \forall j\in\mathbb{I}_1^m.
\end{flalign}
Owing to the property $T(0)=0$, the first-order  virtual controller  $\alpha_{1,j}(t)$ in (\ref{a1}) admits the following
\begin{flalign}
	\alpha_{1,j}(t_{\max}^-)=-k_1%\ln\left(\frac{1+\xi_{1,j}(t_j^k)}{1-\xi_{1,j}(t_j^k)} \right)
	T\left(\xi_{1,j}(t_{\max}^-) \right)\to 0,\ \ \forall j\in\mathbb{I}_1^m.
\end{flalign}
Then, we can further derive from (\ref{xiijt}) that, for all $j\in\mathbb{I}_1^m$,
\begin{flalign}
	\xi_{i,j}(t_{\max}^-)=  \frac{x_{i,j}(t_{\max}^-)-\alpha_{i-1,j}(t_{\max}^-)}{\rho_{i,j}(t_{\max}^-)+\eta_{i,j} \sigma(t_{\max}^-)}\to 0, \ \ i\in\mathbb I_2^n 
\end{flalign}
and thus the $i$th-order  virtual controller $\alpha_{i,j}(t)$ in (\ref{ai}) admits
\begin{flalign}
	\alpha_{i,j}(t_{\max}^-)=-k_i%\ln\left(\frac{1+\xi_{1,j}(t_j^k)}{1-\xi_{1,j}(t_j^k)} \right)
	T\left(\xi_{i,j}(t_{\max}^-) \right)\to 0,\ \ i\in\mathbb I_2^n,\ \ j\in\mathbb{I}_1^m. \label{aijt}
\end{flalign}
As a result, we know from (\ref{aijt}) that
\begin{flalign}
	u_j(t_{\max}^-)=\alpha_{n,j}(t_{\max}^-)\to 0,\ \ \forall j\in\mathbb{I}_1^m. \label{ujtm} 
\end{flalign} 
Owing to (\ref{hs}), (\ref{ujtm}) and the continuity of $u_j(t)$, $\forall j\in\mathbb{I}_1^m$ and $\sigma (t)$, there must exist an unknown time instant  $T\in[0,t_{\max})$ %integer $K_1>0$ 
such that $|u_j(t)|\le\bar u_j$, $ j\in\mathbb{I}_1^m$ and $\sigma (t)>0$, $\forall t\in[T,t_{\max})$.  %$\forall k\ge K_1$. 
Then, we have $|\Delta u_j(t)|=|\operatorname{sat}(u_j(t))-u_j(t)|=0$,  $\forall j\in\mathbb{I}_1^m$ and $\forall t\in[T,t_{\max})$. Moreover, we know from the definition of $\Delta\hat u (t)$ in (\ref{huj}) that $-\hat u \le\Delta\hat u (t)\le 0$,   $\forall t\ge 0$. Therefore, it can be deduced from (\ref{dsigma}) that
\begin{flalign}
	\dot\sigma (t)\le-\beta \sigma (t)<0, \ \ \forall t\in[T,t_{\max}).
\end{flalign}
Hence, $\sigma (t)$ is strictly decreasing, %at those points
$\forall t\in[T,t_{\max})$. \iffalse However, since $\sigma(t_{\max}^-)\to +\infty$, we can extract a %sub-sequence 
sub-interval of $[T_j,t_{\max})$ where $\sigma (t)$ is increasing. % (by choosing points with larger values). 
At such %points
sub-interval, the derivative $\dot\sigma (t)$ cannot be negative,\fi
Then, we have
\begin{flalign}
	\sigma (t)\le\sigma (T),\ \  \forall t\in[T,t_{\max}).
\end{flalign}
This leads to a contradiction  %$\sigma (t)\le\sigma (t_j^{K_1})<+\infty$, $\forall t\in[0,t_{\max})$. This contradicts 
to the hypothesis $\lim_{t\to t_{\max}^-}\sigma(t)=+\infty$ in (\ref{hs}). Therefore, the existence of unknown constant $\bar\sigma >0$ such that $|\sigma(t)|\le \bar\sigma <+\infty$, $\forall t\in[0,t_{\max})$ is guaranteed. %This ends the proof. \hfill$\blacksquare$

Next, we exclude the occurrence of $\lim_{t\to t_{\max}^-}|\xi_{i,j}(t)|=1$, $\forall i\in\mathbb I_1^{n-1}$ and $\forall j\in\mathbb I_1^{m}$. %, as shown in \textcolor{black}{Lemma 4}.
%\textcolor{black}{\textit{Lemma 4:}} For all $i\in\mathbb I_1^{n-1}$ and $j\in\mathbb I_1^{m}$, there must exist  unknown constants $\bar\xi_{i,j}>0$ such that $|\xi_{i,j}(t)|\le\bar\xi_{i,j}<1$, $\forall t\in[0,t_{\max})$.
%\textit{Proof:}  
We assume that there exist some $i\in\mathbb I_i\subset\mathbb I_1^{n-1}$ and $j\in\mathbb I_j\subset\mathbb I_1^{m}$ (where $\mathbb{I}_i\ne \{0\}$ and $\mathbb{I}_j\ne \{0\}$) such that
\begin{flalign}
	\lim_{t\to t_{\max}^-}|\xi_{i,j}(t)|=1. \label{xiij}
\end{flalign}
Then, owing to the property $\lim_{\diamond\to\pm 1}T(\diamond)=\pm\infty$, the $i$th-order virtual controller $\alpha_{i,j}(t)$, $i\in\mathbb I_i$, $j\in\mathbb I_j$ in (\ref{ai}) admits the following
\begin{flalign}
	|\alpha_{i,j}(t_{\max}^-)|=
	|-k_iT\left(\xi_{i,j}(t_{\max}^-) \right)|\to +\infty. \label{aij}
\end{flalign}
From the definition of $\xi_{i,j}(t)$ in (\ref{xi1jt}) and (\ref{xiijt}), we have 
\begin{flalign}
	&\alpha_{i,j}(t)=x_{i+1,j}(t)-\xi_{i+1,j}(t)\left(\rho_{i+1,j}(t)+ \eta_{i+1,j}\sigma (t)\right)\notag\\
	&\ \forall i\in\mathbb I_1^{n-1},\ \ \forall j\in\mathbb I_1^m. \label{aijte}
\end{flalign}
We have shown %in the proof of \textcolor{black}{Lemma 3} 
that, for all $i\in\mathbb{I}_1^{n-1}$ and $j\in\mathbb{I}_1^m$, $x_{i+1,j}(t)$ are bounded, $\forall t\ge 0$. 
From (\ref{uni}), % and \textcolor{black}{Lemma 3}, 
we know that for all $i\in\mathbb{I}_1^{n-1}$ and $j\in\mathbb{I}_1^m$, $|\xi_{i+1,j}(t)|< 1$ and $|\sigma(t)|\le\bar\sigma <+\infty$ hold, $\forall t\in[0,t_{\max})$. By construction, $\rho_{i+1,j}(t)$, $\forall i\in\mathbb{I}_1^{n-1}$, $\forall j\in\mathbb{I}_1^m$ are bounded, $\forall t\ge 0$. Since %$c_{i+1,j}$, $\forall i\in\mathbb{I}_1^{n-1}$, $\forall j\in\mathbb{I}_1^m$ and $\mu$ 
$\eta_{i+1,j}$, $\forall i\in\mathbb{I}_1^{n-1}$, $\forall j\in\mathbb{I}_1^m$ are positive constants, we can conclude from (\ref{aijte}) that, for all $i\in\mathbb I_1^{n-1}$ and $j\in\mathbb I_1^m$, $\alpha_{i,j}(t)$ are bounded, $\forall t\in[0,t_{\max})$. This contradicts to (\ref{aij}). Therefore, the hypothesis in (\ref{xiij}), i.e., $\lim_{t\to t_{\max}^-}|\xi_{i,j}(t)|=1$, $\forall i\in\mathbb I_i$, $\forall j\in\mathbb I_j$ is impossible, and the existence of constants $\bar\xi_{i,j}>0$ such that $|\xi_{i,j}(t)|\le\bar\xi_{i,j}<1$, $\forall i\in\mathbb I_1^{n-1}$, $\forall j\in\mathbb I_1^{m}$, $\forall t\in[0,t_{\max})$ is ensured. %This completes the proof. \hfill$\blacksquare$

Now, we  exclude $\lim_{t\to t_{\max}^-}|\xi_{n,j}(t)|=1$, $\forall j\in\mathbb{I}_1^m$. %, as shown in \textcolor{black}{Lemma 5}.
%\textcolor{black}{\textit{Lemma 5:}} For all $j\in\mathbb I_1^{m}$, there must exist  unknown constants $\bar\xi_{n,j}>0$ such that $|\xi_{n,j}(t)|\le\bar\xi_{i,j}<1$, $\forall t\in[0,t_{\max})$.
%\textit{Proof:} 
We
assume that there exist some $j\in\mathbb I_j\subset\mathbb I_1^{m}$ (where  $\mathbb{I}_j\ne \{0\}$)  %there exists a time instant $t'\in(0,t_{\max}]$ 
such that 
\begin{flalign}
	\lim_{t\to t_{\max}^-}|\xi_{n,j}(t)|=1.\label{h3}
\end{flalign} 
Consider the following barrier function candidate:
\begin{flalign}
	V_{n,j}(t):=1-|\xi_{n,j}(t)|, \ \ j\in\mathbb I_j,\ \ t\in[0,t_{\max}). \label{Vj}
\end{flalign}
Notice that  $V_{n,j}(t)$ is not differentiable at $\xi_{n,j}(t)=0$. %no matter $\lim_{t\to t_{\max}^-}\xi_{n,j}(t)=1$ or $\lim_{t\to t_{\max}^-}\xi_{n,j}(t)=-1$ holds, 
However, from the hypothesis (\ref{h3}) and the continuity of $\xi_{n,j}(t)$, $\forall j\in \mathbb I_1^{m}$, there exist unknown time instants ${T}_j \in[0,t_{\max})$, $j\in\mathbb I_j$ such that
\begin{flalign}
	&\operatorname{sgn}(\xi_{n,j}(t))=  \left\{ \begin{array}{c}
		1,\ \  \text{if}\ \lim_{t\to t_{\max}^-}\xi_{n,j}(t)=1\\
		-1,\ \  \text{if}\ \lim_{t\to t_{\max}^-}\xi_{n,j}(t)=-1\\
	\end{array} \right.\notag\\ 
	&\ \forall j\in\mathbb I_j, \ \ \forall t\in[{T}_j ,t_{\max}). \label{sgnxi}
\end{flalign}
Next, for each $j\in\mathbb I_j$, we discuss these two cases over $[{T}_j ,t_{\max})$.

%\textit{Scenario 1:} $\xi_{n,j}(t)=0$. In such scenario, the cases $\lim_{t\to t_{\max}^-}|\xi_{n,j}(t)|=1$, $\forall j\in\mathbb{I}_1^m$ are excluded.

\textit{\underline{Case 1}:} \textit{For all $t\in[{T}_j ,t_{\max})$ where $j\in\mathbb I_j$,  it holds that $\operatorname{sgn}(\xi_{n,j}(t))=1$ and $\lim_{t\to t_{\max}^-}\xi_{n,j}(t)=1$.} In this case, we have 
\begin{flalign}
	V_{n,j}(t)=1-\xi_{n,j}(t),\ \ \forall t\in[{T}_j ,t_{\max}), \ \ j\in\mathbb I_j.
\end{flalign}
Then, differentiating $V_{n,j}(t)$ along the trajectory (\ref{dxii}) results in
\begin{flalign}
	&\dot V_{n,j}(t)= -\dot\xi_{n,j}(t) =
	-\frac{1}{\rho_{n,j}(t)+\eta_{n,j} \sigma (t)}\Big(f_{n,j}(\tilde x_n(t),t)%+g_{n,j,j}\operatorname{sat}(u_j)\notag\\& 
	\notag\\&  +\sum_{l=1 %(l\ne j)
	}^{m}g_{n,j,l}(\tilde x_n(t),t)
	\times\operatorname{sat}\left(u_l(t) \right)-\dot \alpha_{n-1,j}(t)
	-\xi_{n,j}(t)\dot\rho_{n,j}(t) \notag\\&+ \eta_{n,j}\beta \xi_{n,j}(t)
	 \sigma(t)-c_{n,j}\xi_{n,j}(t)\|\Delta u(t)\|_1 -\eta_{n,j}\gamma \xi_{n,j}(t)\Delta \hat u (t)\Big)\notag\\
	& \  \textcolor{black}{\forall t\in[{T}_j ,t_{\max}), \ \ j\in\mathbb I_j.} \label{dVj}
\end{flalign}
In (\ref{dVj}), $\rho_{n,j}(t)$ and $\dot\rho_{n,j}(t)$, $\forall j\in\mathbb I_1^m$, $\forall t\ge 0$ are bounded by construction; $\eta_{n,j},c_{n,j},\beta ,\gamma >0$, $\forall j\in\mathbb I_1^m$  are constants; $\sigma(t)$, $\forall t\in[0,t_{\max})$ is bounded (before (\ref{h3}));  $\operatorname{sat}\left(u_l(t) \right)$, $\forall l\in\mathbb I_1^m$, $\forall t\ge 0$ and $\Delta \hat u (t)$, $\forall t\ge 0$  are bounded from the definition in (\ref{sat}) and (\ref{huj}), respectively; and $|\xi_{n,j}(t)|< 1$, $\forall j\in\mathbb I_1^m$, $\forall t\in[0,t_{\max})$ has been established in (\ref{uni}). Additionally, $x_{i,j}(t)$, $\forall i\in\mathbb I_1^n$, $\forall j\in\mathbb I_1^m$, $\forall t\ge 0$ are bounded (as shown %in the proof of \textcolor{black}{Lemma 3}
earlier). By using \textcolor{black}{Assumption 2} and applying the Extreme Value Theorem to $\bar f_n(\tilde x_n(t))$ and $\bar g_n(\tilde x_n(t))$, the boundedness of $f_{n,j}(\tilde x_n(t),t)$ and  $g_{n,j,l}(\tilde x_n(t),t)$, $\forall j\in\mathbb I_1^m$, $\forall l\in\mathbb I_1^m$, $\forall t\in[0,t_{\max})$ is ensured.  Therefore, there exist unknown constants $C_{n,j}>0$, $j\in\mathbb I_1^m$ such that
\begin{flalign}
	\Big|&f_{n,j}(\tilde x_n(t),t) +\sum_{l=1 }^{m}g_{n,j,l}(\tilde x_n(t),t)\operatorname{sat}\left(u_l(t) \right)-\xi_{n,j}(t)\notag\\&
	\times\dot\rho_{n,j}(t) +\eta_{n,j}\beta \xi_{n,j}(t)\sigma(t)
	-\eta_{n,j}\gamma \xi_{n,j}(t)\Delta \hat u (t)\Big|\notag\\
	&\le C_{n,j},\ \ \forall j\in\mathbb I_1^m,\ \  \forall t\in[0,t_{\max}). \label{C_j}
\end{flalign}
Moreover, with $c_{n,j}\gamma  \hat u <\mu\beta \rho_{n,j,\infty}$, $j\in\mathbb I_1^m$ in (\ref{cij}), it holds that
\begin{flalign}
	\rho_{n,j}(t)+\eta_{n,j} \sigma (t)%\ge \rho_{n,j,\infty}+\eta_{n,j} \underline\sigma 
	>0,\ \  \forall j\in\mathbb I_1^m,\ \  \forall t\ge 0. \label{rnj}
\end{flalign}
Then, substituting (\ref{C_j})-(\ref{rnj}) into (\ref{dVj}) yields
\begin{flalign}
	\dot V_{n,j}(t)\ge&
	-\frac{C_{n,j}-\dot \alpha_{n-1,j}(t)
		-c_{n,j}\xi_{n,j}(t)\|\Delta u(t)\|_1}{\rho_{n,j}(t)+\eta_{n,j} \sigma (t)}\notag\\
	=&\ \frac{\dot \alpha_{n-1,j}(t)
		+c_{n,j}\xi_{n,j}(t)\|\Delta u(t)\|_1-C_{n,j}}{\rho_{n,j}(t)+\eta_{n,j} \sigma (t)}\notag\\
		&\ \forall t\in[{T}_j ,t_{\max}), \ \ j\in\mathbb I_j. \label{dVjg}
\end{flalign}
Now, we give a rough estimation of the bound for $\dot \alpha_{n-1,j}(t)$ in (\ref{dVjg}). From (\ref{dai}) and (\ref{dxii}), $\dot\alpha_{i,j}(t)$, $i\in\mathbb I_1^{n-1}$, $j\in\mathbb I_1^{m}$ are calculated as
\begin{flalign}
	&\dot\alpha_{i,j}(t)=  -k_i\frac{dT(\xi_{i,j}(t))}{d\xi_{i,j}(t)}\dot\xi_{i,j}(t)
	=\frac{-k_i}{\rho_{i,j}(t)+\eta_{i,j} \sigma (t)}\frac{dT(\xi_{i,j}(t))}{d\xi_{i,j}(t)} \notag\\&\times\Big(f_{i,j}(\tilde x_i(t),t)
	+\sum_{l=1}^{m}g_{i,j,l}(\tilde x_i(t),t)x_{i+1,l}(t)-\dot \alpha_{i-1,j}(t)\notag\\& -\xi_{i,j}(t)
	\times\dot\rho_{i,j}(t) +\eta_{i,j}\beta \xi_{i,j}(t)\sigma(t)-c_{i,j}\xi_{i,j}(t)\|\Delta u(t)\|_1\notag\\&
	-\eta_{i,j}\gamma\xi_{i,j}(t)\Delta \hat u (t)\Big), \ \ \forall t\in[0,t_{\max}). \label{daij}
\end{flalign}
\iffalse
\begin{flalign}
	\dot\alpha_{i,j}(t)=& -k_i\frac{dT(\xi_{i,j}(t))}{d\xi_{i,j}(t)}\dot\xi_{i,j}(t)\notag\\
	=&-k_i\frac{dT(\xi_{i,j}(t))}{d\xi_{i,j}(t)} \frac{1}{\rho_{i,j}(t)+\eta_{i,j} \sigma (t)}\Big(f_{i,j}(\tilde x_i(t),t)\notag\\&
	+\sum_{l=1}^{m}g_{i,j,l}(\tilde x_i(t),t)x_{i+1,l}(t)-\dot \alpha_{i-1,j}(t)-\xi_{i,j}(t)\notag\\&
	\times\dot\rho_{i,j}(t) +\eta_{i,j}\beta \xi_{i,j}(t)\sigma(t)-c_{i,j}\xi_{i,j}(t)\|\Delta u(t)\|_1\notag\\&
	-\eta_{i,j}\gamma\xi_{i,j}(t)\Delta \hat u (t)\Big), \ \ \forall t\in[0,t_{\max}). \label{daij}
\end{flalign}\fi 
%By \textcolor{black}{Lemma 4}, and 
According to the properties $0<[dT(\diamond)/d\diamond]<+\infty$ and $\lim_{\diamond\rightarrow \pm1} [dT(\diamond)/d\diamond]=+\infty$, we know that 
\begin{flalign}
	0&<\frac{dT(\xi_{i,j}(t))}{d\xi_{i,j}(t)}\le \frac{dT\big(\xi_{i,j}(t)\big)}{d\xi_{i,j}(t)}|_{\xi_{i,j}(t)=\bar\xi_{i,j}}:=D_{i,j}<+\infty\notag\\
	&\forall i\in\mathbb I_1^{n-1},\ \ \forall j\in\mathbb I_1^{m},\ \ \forall t\in[0,t_{\max}) \label{dTxiij}
\end{flalign}
where $D_{i,j}>0$ are unknown constants. With the gain selection $c_{i,j}\gamma  \hat u <\mu\beta \rho_{i,j,\infty}$, $i\in\mathbb I_1^{n-1}$, $j\in\mathbb I_1^{m}$ in (\ref{cij}), it holds that
\begin{flalign}
	\rho_{i,j}(t)+\eta_{i,j} \sigma (t)\ge \rho_{i,j,\infty}+\eta_{i,j} \underline\sigma >0,\ \  \forall t\ge 0. \label{rij}
\end{flalign}
Following a  reasoning similar to that after (\ref{dVj}), we know that, for all $i\in\mathbb I_1^{n-1}$, $j\in\mathbb I_1^{m}$ and $l\in\mathbb I_1^{m}$, the terms $f_{i,j}(\tilde x_i(t),t)$,  $g_{i,j,l}(\tilde x_i(t),t)$, $x_{i+1,l}(t)$, $\dot\rho_{i,j}(t)$, $\sigma(t)$ and $\Delta \hat u (t)$  are bounded, $\forall t\in[0,t_{\max})$. Moreover, $\eta_{i,j}, c_{i,j}, \beta ,\gamma$, $i\in\mathbb I_1^{n-1}$, $j\in\mathbb I_1^{m}$ are positive constants. Since $|\xi_{i,j}(t)|\le\bar\xi_{i,j}< 1$, $\forall i\in\mathbb I_1^{n-1}$, $\forall j\in\mathbb I_1^{m}$, $\forall t\in[0,t_{\max})$, there  exist unknown constants $C_{i,j}>0$ such that 
\begin{flalign}
	\Big|&f_{i,j}(\tilde x_i(t),t) +\sum_{l=1 }^{m}g_{i,j,l}(\tilde x_i(t),t)x_{i+1,l}(t)-\xi_{i,j}(t)\dot\rho_{i,j}(t) \notag\\&+\eta_{i,j}\beta \xi_{i,j}(t)\sigma(t)
	-\eta_{i,j}\gamma\xi_{i,j}(t)\Delta \hat u (t)\Big|\le C_{i,j}\notag\\
	&\ \forall i\in\mathbb I_1^{n-1},\ \ \forall j\in\mathbb I_1^{m},\ \ \forall t\in[0,t_{\max}). \label{Cij}
\end{flalign}
Then, inserting %(\ref{dTxiij}), (\ref{rij}) and 
(\ref{dTxiij})-(\ref{Cij}) into (\ref{daij}) yields
\begin{flalign}
	|\dot\alpha_{i,j}(t)|\le&\ \frac{k_i D_{i,j}\left( |\dot\alpha_{i-1,j}(t)|+c_{i,j}\bar\xi_{i,j}\|\Delta u(t)\|_1+C_{i,j} \right)}{\rho_{i,j,\infty}+\eta_{i,j} \underline\sigma }\notag\\
	=&\ A_{i,j}|\dot\alpha_{i-1,j}(t)|+B_{i,j}\|\Delta u(t)\|_1+ C_{i,j}'\notag\\
	 \forall& i\in\mathbb I_1^{n-1},\ \ \forall j\in\mathbb I_1^{m},\ \ \forall t\in[0,t_{\max}). \label{daijl}
\end{flalign}
where $A_{i,j}:= \frac{k_i D_{i,j}}{\rho_{i,j,\infty}+\eta_{i,j} \underline\sigma }$, $B_{i,j}:=A_{i,j}c_{i,j}\bar\xi_{i,j}$ and $C_{i,j}':=A_{i,j}C_{i,j}$ are unknown positive constants. Using (\ref{daijl}), the bound of $\dot\alpha_{n-1,j}(t)$, $j\in\mathbb I_j$, $t\in[{T}_j ,t_{\max})$ can be obtained via a recursive analysis. Letting $i=1$, it follows from (\ref{daijl}) and \textcolor{black}{Assumption 3} that
\begin{flalign}
	|\dot\alpha_{1,j}(t)|\le&\ A_{1,j}|\textcolor{black}{\dot\alpha_{0,j}(t)}|+B_{1,j}\|\Delta u(t)\|_1+ C_{1,j}'\label{da1jl} \\
	\le&\ A_{1,j}\bar {\dot y}_{d,j}+B_{1,j}\|\Delta u(t)\|_1+ C_{1,j}'   \notag\\
	=&\ K_{1,j}\|\Delta u(t)\|_1+L_{1,j}
	,\ \ \forall j\in\mathbb I_1^{m},\ \ \forall t\in[0,t_{\max}) \notag
\end{flalign}
where $K_{1,j}:=B_{1,j}$ and $L_{1,j}:=A_{1,j}\bar {\dot{y}}_{d,j}+C_{1,j}'$ are unknown positive constants.
Further, letting $i=2$, we have 
\begin{flalign}
	|\dot\alpha_{2,j}(t)|\le &\ A_{2,j}|\dot\alpha_{1,j}(t)|+B_{2,j}\|\Delta u(t)\|_1+ C_{2,j}'\label{da2jl} \\
	\le &\  A_{2,j}(K_{1,j}\|\Delta u(t)\|_1+L_{1,j})+B_{2,j}\|\Delta u(t)\|_1+ C_{2,j}' \notag\\
	= &\ K_{2,j}\|\Delta u(t)\|_1+L_{2,j}
	,\ \ \forall j\in\mathbb I_1^{m},\ \ \forall t\in[0,t_{\max}) \notag
\end{flalign}
where $K_{2,j}:=A_{2,j}K_{1,j}+B_{2,j}$ and $L_{2,j}:=A_{2,j} L_{1,j}+C_{2,j}'$ are unknown positive constants. Similarly and recursively, we have
\begin{flalign}
	&|\dot\alpha_{i,j}(t)|\le K_{i,j}\|\Delta u(t)\|_1+L_{i,j}\notag\\
	&\ \forall i\in\mathbb I_1^{n-1},\ \ \forall j\in\mathbb I_1^{m},\ \ \forall t\in[0,t_{\max}) \label{daijle}
\end{flalign}
with \textcolor{black}{$K_{i,j}:= \sum_{p=1}^{i}\left(B_{p,j}\prod_{q=p+1}^{i}A_{q,j} \right)$ and $L_{i,j}:=\bar {\dot y}_{d,j}\prod_{p=1}^{i}A_{p,j}+ \sum_{p=1}^{i}\left(C'_{p,j}\prod_{q=p+1}^{i}A_{q,j} \right)$} unknown positive constants with $\prod_{q=i+1}^{i}A_{q,j}:=1$. Hence, $\dot\alpha_{n-1,j}(t)$'s bound is
\begin{flalign}
	&|\dot\alpha_{n-1,j}(t)|\le K_{n-1,j}\|\Delta u(t)\|_1+L_{n-1,j}\notag\\
	&\  \forall j\in\mathbb I_1^{m},\ \ \forall t\in[0,t_{\max}). \label{dan-1jl}
\end{flalign}
%where $K_{n-1,j}:= \sum_{i=1}^{n-1}\left(B_{i,j}\prod_{l=i+1}^{n-1}A_{l,j} \right)$ and $L_{n-1,j}:=\bar y_{d,j}\prod_{l=1}^{n-1}A_{l,j}+ \sum_{i=1}^{n-1}\left(C'_{i,j}\prod_{l=i+1}^{n-1}A_{l,j} \right)$.
Substituting (\ref{dan-1jl}) into (\ref{dVjg}) results in
\begin{flalign}
	\dot V_{n,j}(t)\ge&\ \frac{1}{\rho_{n,j}(t)+\eta_{n,j} \sigma (t)}(-K_{n-1,j}\|\Delta u(t)\|_1 -L_{n-1,j}
	\notag\\&\ +c_{n,j}\xi_{n,j}(t)\|\Delta u(t)\|_1-C_{n,j})\notag\\
		=&\ \frac{\left(c_{n,j}\xi_{n,j}(t)-K_{n-1,j} \right)\|\Delta u(t)\|_1-L_{n-1,j}
			-C_{n,j}}{\rho_{n,j}(t)+\eta_{n,j} \sigma (t)}
		\notag\\   \forall& t\in[{T}_j ,t_{\max}), \ \ j\in\mathbb I_j. \label{dVjge}
\end{flalign}
To proceed, we decompose %$\|\Delta u(t)\|_1$ as 
$\|\Delta u(t)\|_1=\sum_{j\in\mathbb I_j}|\Delta u_j(t)|+\sum_{j\in\mathbb I_1^m\setminus \mathbb I_j}|\Delta u_j(t)|$. %For all $j\in \mathbb I_1^m\setminus \mathbb I_j$, form the definition of $\Delta u_j(t)$, $\operatorname{sat}(u_j(t))$ in (\ref{dujt}) and (\ref{sat}), the fact  $\lim_{t\to t_{\max}^-}|\xi_{n,j}(t)|<1$, (\ref{ai}), and the property $\lim_{\diamond\to \pm1}T(\diamond)=\pm\infty$,  we know that $\sum_{j\in\mathbb I_1^m\setminus \mathbb I_j}|\Delta u_j(t)|$ is bounded, $\forall t\in[0,t_{\max})$. Moreover, 
Recall $\lim_{t\to t_{\max}^-}|\xi_{n,j}(t)|=1$, $\forall j\in\mathbb I_j$ under the hypothesis (\ref{h3}). From the design of $u(t)$ in (\ref{ai}) and the property %$\lim_{\xi_{n,j}(t)\to 1}T(\xi_{n,j}(t))=+\infty$
$\lim_{\diamond\to \pm1}T(\diamond)=\pm\infty$, we know that
\begin{flalign}
	|u_j(t_{\max}^-)|=|-k_nT(\xi_{n,j}(t_{\max}^-))|\to +\infty,\ \ \forall j\in\mathbb I_j. \label{uj}
\end{flalign}
Owing to (\ref{uj}) and the continuity of $u_{j}(t)$, $\forall j\in\mathbb I_1^m$, there must exist unknown time instants $T'_j  \in[{T}_j ,t_{\max})$ such that
\begin{flalign}
	|u_j(t)|\ge \bar u_j,\ \ \forall t\in[T'_j  ,t_{\max}),\ \ \forall j\in\mathbb I_j.
\end{flalign}
Then, it follows from the definition of $\Delta u_j(t)$ in (\ref{dujt}) that
\begin{flalign}
	|\Delta u_j(t)|=&\ |\operatorname{sat}(u_j(t))-u_j(t)|=|\bar u_j\operatorname{sgn}(u_j(t))-u_j(t)|\notag\\
	=&\ \left\{ \begin{array}{l}
		|\bar u_j+u_j(t)|,\ \ \text{if}\ u_{j}(t)\le -\bar u_{j}\\
		|\bar u_j-u_j(t)|,\ \ \text{if}\ u_{j}(t)\ge \bar u_{j}\\
	\end{array} \right.\notag\\
	=&\ |u_j(t)|-\bar u_j,\ \  \forall t\in[T'_j  ,t_{\max}), \ \ \forall j\in\mathbb I_j.\label{duj}
\end{flalign}
Leveraging (\ref{duj}), (\ref{dVjge}) becomes
\begin{flalign}
	\dot V_{n,j}(t)\ge&\ \frac{1}{\rho_{n,j}(t)+\eta_{n,j} \sigma (t)} ((  c_{n,j}\xi_{n,j}(t)-K_{n-1,j} )\label{dVjge1} \\
	& \times ( \sum_{j\in\mathbb I_j}\left( |u_j(t)|-\bar u_j \right)+\sum_{j\in\mathbb I_1^m\setminus \mathbb I_j}|\Delta u_j(t)| )\notag\\&
	-L_{n-1,j}
		-C_{n,j}),\
	 \  \forall t\in[T'_j  ,t_{\max}),\ \ j\in\mathbb I_j. \notag
\end{flalign}
From the definition of unknown constants $K_{n-1,j}>0$, $j\in\mathbb I_1^m$ after (\ref{daijle}), we know that $K_{n-1,j}$ are \textcolor{black}{independent} of $c_{n,j}$. Since $c_{n,j}>K_{n-1,j}$ by (\ref{c2}), it holds that
\begin{flalign}
	0<\frac{c_{n,j}+K_{n-1,j}}{2c_{n,j}}<1,\ \ \forall j\in\mathbb I_1^m.
\end{flalign}
Then, owing to the continuity of $\xi_{n,j}(t)$, $\forall j\in\mathbb I_1^m$ and the hypothesis $\lim_{t\to t_{\max}^-}\xi_{n,j}(t)=1$,  $j\in\mathbb I_j$ under Scenario 1, there exist unknown time instants $T''_j  \in[T'_j  ,t_{\max})$ such that
\begin{flalign}
	\xi_{n,j}(t)>\frac{c_{n,j}+K_{n-1,j}}{2c_{n,j}},\ \  \forall t\in[T''_j  ,t_{\max}),\ \ j\in\mathbb I_j.
\end{flalign}
Therefore, we have
\begin{flalign}
	&c_{n,j}\xi_{n,j}(t)-K_{n-1,j}>\frac{c_{n,j}-K_{n-1,j}}{2}>0\notag\\
	&\ \forall t\in[T''_j  ,t_{\max}),\ \ j\in\mathbb I_j. \label{cnj}
\end{flalign}
Note that $c_{n,j},K_{n-1,j},\bar u_j,L_{n-1,j},C_{n,j}>0$, $j\in\mathbb I_1^m$ are constants. By (\ref{uj}), there exist unknown time instants $T'''_j  \in[T''_j  ,t_{\max})$ so that
\begin{flalign}
\sum_{j\in\mathbb I_j}|&u_j(t)|>  \frac{2\left( L_{n-1,j}+C_{n,j} \right) }{c_{n,j}-K_{n-1,j}}+\sum_{j\in\mathbb I_j}\bar u_j\notag\\
	&\ \forall t\in[T'''_j  ,t_{\max}),\ \ j\in\mathbb I_j. \label{uja}
\end{flalign}
By using (\ref{rnj}), (\ref{cnj}), (\ref{uja}); the facts $|\xi_{n,j}(t)|<1$, $\forall j\in\mathbb I_1^m$, $\forall t\in[0,t_{\max})$; $\xi_{n,j}(t)>0$, $\forall t\in[{T}_j ,t_{\max})$, $ j\in\mathbb I_j$; and $\sum_{j\in\mathbb I_1^m\setminus \mathbb I_j}|\Delta u_j(t)|\ge0$, $\forall t\in[0,t_{\max})$, (\ref{dVjge1}) is bounded as 
\begin{flalign}
	\dot V_{n,j}(t)\ge&\ \frac{1}{\rho_{n,j}(t)+\eta_{n,j} \sigma (t)} (\frac{c_{n,j}-K_{n-1,j}}{2} \sum_{j\in\mathbb I_j}\big(|u_j(t)|\notag\\&
	-\bar u_j\big)-L_{n-1,j}
	-C_{n,j}) %\notag\\ \ge&\ \frac{1}{\rho_{n,j}(t)+\eta_{n,j} \sigma (t)}c_{n,j}\left( 1-\xi_{n,j}(t)\right)\bar u_j 
	\notag\\
	>&\ 0,\ \ \forall t\in[T'''_j  ,t_{\max}),\ \ j\in\mathbb I_j \label{dVjge2}
\end{flalign}
which implies that $V_{n,j}(t)$, $j\in\mathbb I_j$ is strictly increasing, 
$\forall t\in[T'''_j  ,t_{\max})$. As a result, we have
\begin{flalign}
	&V_{n,j}(t)\ge  V_{n,j}(T'''_j  )=1-\xi_{n,j}(T'''_j  )>0\notag\\
	&\  \forall t\in[T'''_j  ,t_{\max}),\ \ j\in\mathbb I_j.
\end{flalign}
This leads to a contradiction  
to the hypothesis $\lim_{t\to t_{\max}^-}\xi_{n,j}(t)=1$, $j\in\mathbb I_j$, at which $\lim_{t\to t_{\max}^-}V_{j}(t)=0$, $j\in\mathbb I_j$. Therefore, the occurrence of $\lim_{t\to t_{\max}^-}\xi_{n,j}(t)=1$, $\forall j\in\mathbb I_1^m$ is impossible.

\textit{\underline{Case 2}:} \textit{For all $t\in[{T}_j ,t_{\max})$ where $j\in\mathbb I_j$,  it holds that $\operatorname{sgn}(\xi_{n,j}(t))=-1$ and $\lim_{t\to t_{\max}^-}\xi_{n,j}(t)=-1$.} In this case, by following a  reasoning similar to Case 1, we can also obtain that the occurrence of $\lim_{t\to t_{\max}^-}\xi_{n,j}(t)=-1$, $\forall j\in\mathbb I_1^m$ is impossible.

In conclusion, we have shown that  the hypothesis (\ref{h3}), i.e., $\lim_{t\to t_{\max}^-}|\xi_{n,j}(t)|=1$, $\forall j\in\mathbb I_j$ is not true under both cases. %\textcolor{black}{That is to say, the case $t_{\max}<+\infty$ does not attribute to the occurrence of $\lim_{t\to t_{\max}^-}|\xi_{n,j}(t)|=1$.}
That is to say, the existence of unknown constants $\bar\xi_{n,j}>0$ such that $|\xi_{n,j}(t)|\le \bar\xi_{n,j}<1$, $\forall j\in\mathbb I_1^m$, $\forall t\in[0,t_{\max})$ is guaranteed.
%This completes the proof. \hfill$\blacksquare$

%With the establishment of \textcolor{black}{Lemmas 3-5}, we can conclude that, 
Now, $\forall t\in[0,t_{\max})$, the unique and maximal solution $z(t)$ for (\ref{dz}) remains strictly within the compact subset  $\prod_{i=1}^{n}\prod_{j=1}^{m}[ -\bar\xi_{i,j},\bar\xi_{i,j}  ]
\times [ -\bar\sigma,\bar\sigma]:=\Omega'_z$.
\iffalse
\begin{flalign}
	\prod_{i=1}^{n}\prod_{j=1}^{m}[ -\bar\xi_{i,j},\bar\xi_{i,j}  ]
	\times [ -\bar\sigma,\bar\sigma  ]:=\Omega'_z.\label{Om}
\end{flalign}
\fi 
That is, $z(t)\in\Omega'_z\subset\Omega_z$, $\forall t\in[0,t_{\max})$. Henceforth, what remains to prove in Stage II is  the boundedness of all closed-loop signals in (\ref{dz}), $\forall t\in[0,t_{\max})$.
This is done by using the following: 1) the boundedness of terms $x_{i,j}(t)$,  $f_{i,j}(\tilde x_{i}(t),t)$, $g_{i,j,l}(\tilde x_{i}(t),t)$, $\operatorname{sat}\left(u_l(t) \right)$ and $\sigma (t)$, $\forall i\in\mathbb I_1^n$, $\forall j\in\mathbb I_1^m$, $\forall l\in\mathbb I_1^m$, $\forall t\in[0,t_{\max})$; %, as shown in the proof of \textcolor{black}{Lemmas 3 and 5}; 
 2) the fact $|\xi_{i,j}(t)|\le\bar\xi_{i,j}<1$, $\forall i\in\mathbb I_1^n$, $\forall j\in\mathbb I_1^m$, $\forall t\in[0,t_{\max})$; %, as shown in the proof of \textcolor{black}{Lemmas 4 and 5}; 
3) the properties $\lim_{\diamond\to\pm 1}T(\diamond)=\pm\infty$ and $0<[dT(\diamond)/d\diamond]<+\infty$, which guarantees the boundedness of mapped errors $\epsilon_{i,j}(t)$ in (\ref{ep1j}), (\ref{epij}) and controllers $\alpha_{i,j}(t)$ in (\ref{a1}), (\ref{ai}), $\forall i\in\mathbb I_1^n$, $\forall j\in\mathbb I_1^m$, $\forall t\in[0,t_{\max})$.

\iffalse
Then, it holds from (\ref{ai}) and the property $\lim_{\xi_{n,j}\to\pm 1}[T(\xi_{n,j})]=\pm \infty$ that
\begin{flalign}
	\lim_{t\to t_{\max}^-}|u_j(t)|=\lim_{t\to t_{\max}^-}k_n|T(\xi_{n,j}(t))|=+\infty.
\end{flalign}
From the definition of in , we know that $|\Delta u_j(t)|=|\operatorname{sat}(u_j(t))-u_j(t)|=0$

Then, it can be derived that $\lim_{t\to t_{\max}^-}\sigma (t)$ must be bounded. Otherwise, if $\lim_{t\to t_{\max}^-}|\sigma (t)|=+\infty$, we have that $\lim_{t\to t_{\max}^-}\xi_{1,j}(t)=0$ owing to the boundedness of $x_{1,j}$, $y_{d,j}$ and $\rho_{1,j}$, and thus $\lim_{t\to t_{\max}^-}\alpha_{1,j}(t)=0$ from (\ref{a1}) and the properties $T(0)=0$ and $\lim_{\xi_{1,j}\to\pm 1}[d T(\xi_{1,j})/d \xi_{1,j}]=+\infty$. Similarly and recursively, if $\lim_{t\to t_{\max}^-}|\sigma (t)|=+\infty$, we can obtain that $\lim_{t\to t_{\max}^-}\xi_{i,j}(t)=0$, $i\in\mathbb{I}_2^n$ owing to the boundedness of $x_{i,j}$, $\lim_{t\to t_{\max}^-}\alpha_{i-1,j}(t)$ and $\rho_{i,j}$, and thus $\lim_{t\to t_{\max}^-}\alpha_{i,j}(t)=0$ from (\ref{ai})-(\ref{ai}) and the properties $T(0)=0$ and $\lim_{\xi_{i,j}\to\pm 1}[d T(\xi_{i,j})/d \xi_{i,j}]=+\infty$. This contradicts to the hypothesis $\lim_{t\to t_{\max}^-}|\xi_{n,j}(t)|=1$ since $\lim_{t\to t_{\max}^-}\xi_{n,j}(t)=0$ if $\lim_{t\to t_{\max}^-}|\sigma (t)|=+\infty$. 
 \hfill$\blacksquare$ \fi %Owing to the continuity of $u_j(t)$, there must exist a time instant $\mathcal T\in(0,t')$ such that $u_j(t)$ is bounded, $\forall t\in[0,\mathcal{T}]$, i.e., $|u_j(t)|\le U_j$, $\forall t\in[0,\mathcal{T}]$, where $U_j>0$ is an unknown constant. Then, $\sigma$. 

\textbf{Stage III:} %\textit{extension of the unique and maximal solution of (\ref{dz}) to $z:[0,+\infty)\to\Omega_z$.} 
We have shown in Stage II that $z(t)\in\Omega'_z\subset\Omega_z$, $\forall t\in[0,t_{\max})$, where $\Omega'_z$ is non-empty and compact. By \textcolor{black}{Proposition C.3.6 in \cite{sontag2013mathematical}}, if $t_{\max}<+\infty$, then there exists a time instant $t'\in[0,t_{\max})$ such that $z(t')\notin\Omega'_z$. This contradicts to the fact $z(t)\in\Omega'_z\subset\Omega_z$, $\forall t\in[0,t_{\max})$. Hence, it holds that $t_{\max}=+\infty$. Then, the unique and maximal solution of (\ref{dz}) is extended to $z:[0,+\infty)\to\Omega_z$.

Now, all the closed-loop signals \textcolor{black}{in $z(t)$-dynamics (\ref{dz})} are bounded, $\forall t\ge 0$. Moreover, %it can be seen from the design of modification signal $\sigma(t)$'s dynamics in (\ref{dsigma}) that $\sigma(t)$ is generated by a first-order  dynamics. Further, the control design in Section \ref{sCD} does not involve prior knowledge of the saturation level $\bar u_j$, any feasibility condition, or derivative of the desired tracking trajectories $y_d(t)$. Finally, 
from the definition of $\xi_{1,j}(t)$, $j\in\mathbb I_1^m$ in (\ref{xi1jt}) and result $|\xi_{1,j}(t)|<1$, $\forall j\in\mathbb I_1^m$, $\forall t\ge 0$ in (\ref{uni}), we have
\begin{flalign}
	|e_{j}(t)|<\textcolor{black}{\rho_{1,j}(t)+\eta_{1,j} \sigma(t)},\ \ \forall j\in\mathbb I_1^m,\ \ \forall t\ge 0.
\end{flalign}
Hence, %the satisfaction of modified %output tracking control 
the modified constraints (\ref{PC}) is ensured.
It ends the proof. 
\hfill$\blacksquare$ 

\textcolor{black}{\textit{Remark 5:}} While $K_{n-1,j}>0$, $\forall j\in\mathbb I_1^m$, are unknown constants,  the design parameters $c_{n,j}$ can always be chosen sufficiently large so that the parameter selection condition (\ref{c2}) is met. This principle can also be found in the singular perturbation theory and high-gain observer design \cite{khalil2014high}. In practice, as demonstrated  in Section \ref{sec:simulation}, the simulation performs satisfactorily without selecting excessively large $c_{n,j}$, which indicates the low conservatism of  condition (\ref{c2}).

\textit{Remark 6:} Proving Theorem~1 via conventional 
Lyapunov theory is challenging. To see this, consider the  energy function $V_i(t)=\frac{1}{2}\epsilon_i^\top(t)\epsilon_i(t)$, $i\in\mathbb I_1^n$. Its time derivative $\dot V_i(t)$ involves $\dot \alpha_{i-1}(t)$.
%$\epsilon_i^\top(t)\operatorname{diag}\left(  \frac{dT(\xi_{i,1}(t))}{d\xi_{i,1}(t)},...,\frac{dT(\xi_{i,m}(t))}{d\xi_{i,m}(t)}  \right)\big(g_i(\tilde x_i(t),t)\alpha_i(t)-%\xi_1^*(t)\eta_1\dot\sigma(t)
%\dot \alpha_{i-1}(t)\big)$.
%To address $g_i(\tilde x_i(t),t)\alpha_i(t)$, a controllability condition on $g_i(\tilde x_i(t),t)$ is required, which would restrict the class of admissible system (\ref{sys}). More importantly, 
From (\ref{daij}), the boundedness of $\dot \alpha_{i-1}(t)$, $i\in\mathbb I_2^n$ depends on that of $\|\Delta u(t)\|_1$, which cannot be established at the $i$-th step.

\textcolor{black}{\textit{Remark 7:} In the ISpS case of \cite{10273607}, the gain matrices are diagonal, the saturation levels are identical across all inputs, and the modification dynamics are high-order, multidimensional and non-bidirectional. Our work overcomes  these limitations, albeit at the cost of a more involved parameter selection to satisfy conditions (\ref{c1})-(\ref{c2}).}
It is of practical interest to characterize the bounds on   modification term $\eta_{1,j} \sigma(t)$, $\forall j\in\mathbb I_1^m$, or at least to understand how these bounds can be shaped by design parameters. Next, we derive an exact lower bound and an estimate of the upper bound for the modification signal $\sigma(t)$. These results are summarized in the following theorem.

\textit{Theorem 2:} Consider %the high-order uncertain  highly-coupled MIMO  strict-feedback nonlinear 
system (\ref{sys}) subject to %both 
input saturation  (\ref{sat}) and %output tracking 
performance   constraints (\ref{PC}). %, with the control scheme %developed 
%in Section \ref{sec:results} applied. 
Let all conditions in Theorem 1 hold. Then, for all $j\in\mathbb I_1^m$ and $t\ge 0$, 
it holds that
\begin{flalign}
	-\frac{c_{1,j}\gamma \hat u }{\mu\beta }\le\eta_{1,j} \sigma (t)\le \frac{c_{1,j}}{\beta }\sup_{0\le\tau\le t}\|\Delta u(\tau)\|_1.
\end{flalign}

\textit{Proof:} First, we derive the lower bound of $\eta_{1,j} \sigma (t)$, $j\in\mathbb I_1^m$. To this end, from (\ref{usj}), we have
\begin{flalign}
	\eta_{1,j} \sigma(t)\ge\eta_{1,j} \underline\sigma=-\frac{c_{1,j}\gamma \hat u }{\mu\beta },\ \ \forall j\in\mathbb I_1^m,\ \ \forall t\ge 0.
\end{flalign}
To proceed, we correlate the upper bound of $\eta_{1,j} \sigma (t)$, $j\in\mathbb I_1^m$ with the term $\|\Delta u(t)\|_1$. From (\ref{dhu}), we know that $-\hat u \le\Delta\hat u (t)\le 0$, $\forall t\ge 0$. Therefore, (\ref{dsigma}) is bounded as
\begin{flalign}
	\dot \sigma (t)\le -\beta  \sigma (t) + \mu \|\Delta u(t)\|_1,\ \ \forall t\ge 0. \label{bsj}
\end{flalign}
Then, solving (\ref{bsj}) and using the comparison lemma yields
\begin{flalign}
	\sigma(t)&\le \exp^{-\beta t}\sigma(0)+\exp^{-\beta t}\int_{0}^{t}\exp^{\beta \tau}\mu \|\Delta u(\tau)\|_1d\tau\notag\\
	&\le \mu \exp^{-\beta t}\int_{0}^{t}\exp^{\beta \tau} \sup_{0\le\tau\le t}\|\Delta u(\tau)\|_1d\tau\notag\\
	&=  \mu\sup_{0\le\tau\le t}\|\Delta u(\tau)\|_1 \exp^{-\beta t}\int_{0}^{t}\exp^{\beta \tau} d\tau\notag\\
	&= \frac{\mu}{\beta }\sup_{0\le\tau\le t}\|\Delta u(\tau)\|_1\left(1- \exp^{-\beta t} \right)\notag\\
	&\le \frac{\mu}{\beta }\sup_{0\le\tau\le t}\|\Delta u(\tau)\|_1,\ \ \forall t\ge 0.
\end{flalign}
Therefore, the upper bound of $\eta_{1,j} \sigma (t)$, $j\in\mathbb I_1^m$ is
\begin{flalign}
	\eta_{1,j} \sigma (t)\le \frac{c_{1,j}}{\beta }\sup_{0\le\tau\le t}\|\Delta u(\tau)\|_1,\ \ \forall j\in\mathbb I_1^m,\ \ \forall t\ge 0.
\end{flalign}
This ends the proof.
\hfill$\blacksquare$
\iffalse
Owing to the definition of $\operatorname{sat}\left(u_j(t) \right)$ in (\ref{sat}), we know that, for all $j\in\mathbb I_1^m$ and $t\ge 0$, it holds that
\begin{flalign}
	|\Delta u_j(t)|=|\operatorname{sat}\left(u_j(t) \right)-u_j(t)|\le|u_j(t)|
\end{flalign}\fi

\textcolor{black}{\textit{Remark 8:}} Theorem 2 guides how to adjust the modified performance specifications via design parameters. Specifically, the performance degradation caused by modification term $\eta_{1,j} \sigma(t)$, $j\in\mathbb I_1^m$ can be reduced by setting $\frac{c_{1,j}}{\beta}$ small, while the potential for performance %recovery and 
enhancement %under generally mild saturation 
can be strengthened by setting $\frac{\gamma\hat u}{\mu}$ large. %Based on these insights, 
Therefore, a feasible parameter selection procedure is: 
1) Initially set $\frac{\gamma\hat u}{\mu} = 0$ so that condition (\ref{c1}) is met;
2) Given the desired steady-state accuracy $\rho_{1,j,\infty}$, select $c_{1,j},\beta$ to appropriate values; 
3) Choose $c_{i,j},\rho_{i,j,\infty}$, $i\in\mathbb I_2^{n-1}$ to appropriate values. Note that the permissible upper bound for $\frac{\gamma\hat u}{\mu}$, i.e., $\frac{\beta\rho_{i,j,\infty}}{c_{i,j}}$, must not be less than $\frac{\beta\rho_{1,j,\infty}}{c_{1,j}}$;
4) Set $\rho_{n,j,\infty}$ sufficiently large to provide enough margin for $c_{n,j}$ to satisfy condition (\ref{c2}), and then increase $c_{n,j}$ gradually from a small value to an appropriate one; and
5) Finally, increase $\frac{\gamma\hat u}{\mu}$ from zero to an appropriate value within the allowable range $\frac{\gamma\hat u}{\mu} < \frac{\beta\rho_{1,j,\infty}}{c_{1,j}}$, where  $\hat u$ should be biased  to facilitate activation of the performance enhancement term $\Delta \hat u(t)$ in (\ref{dsigma}). %, where $\hat u$ should be biased to ensure activation of performance enhancement term $\Delta \hat u(t)$ in (\ref{dsigma}).

%\section{Application to Heterogeneous Mobile Robots}\label{sec:application}

%\section{Experiments}\label{sec:experiment}

\section{Simulations}
\label{sec:simulation}
\textcolor{black}{In this section, we make a comparative simulation %between our developed low-complexity input-saturated PPC scheme and 
with the method in \cite{10273607}. Since the gain matrices of %the controlled plant considered 
plants in \cite{10273607} are diagonal, we adopt the following numerical example:
\begin{flalign}
	\dot x_1=x_2,\ \ \dot x_2=-x_2-\begin{bmatrix}
		2x_{1,1}-x_{1,2}+x_{1,1}^3+\sin t\\
		2x_{1,2}-x_{1,1}+x_{1,2}^3+\cos t
	\end{bmatrix}+\operatorname{sat}(u) \notag
	 %\label{msd}
\end{flalign}
where the system nonlinearities are smooth in their arguments.} %\footnote{Since the dimension of mass-spring-damper system (\ref{msd}) is $m=1$, to facilitate presentation, the subscript $j=1$ is omitted throughout Section \ref{PE}.}. 
\iffalse
Obviously, system (\ref{eqf}) is  a special case of system (\ref{sys}) with % $f_1(\tilde x_1(t),t)=0$, $g_1(\tilde x_1(t),t)=1$, $f_2(\tilde x_2(t),t)=-(cx_2(t)+kx_1(t))/{m}$ and $g_2(\tilde x_2(t),t)={m}^{-1}$.
\begin{flalign}
	f_1(x_1(t),t)&=0,\ \ g_1(x_1(t),t)=1\notag\\
	f_2(\tilde x_2(t),t)&=-\frac{kx_1(t)+cx_2(t)}{m},\ \ g_2(\tilde x_2(t),t)=\frac{1}{m} \label{f1}
\end{flalign}
which \fi 
\begin{figure*}[htbp]
	\centering
	{\includegraphics[width=0.46\linewidth]{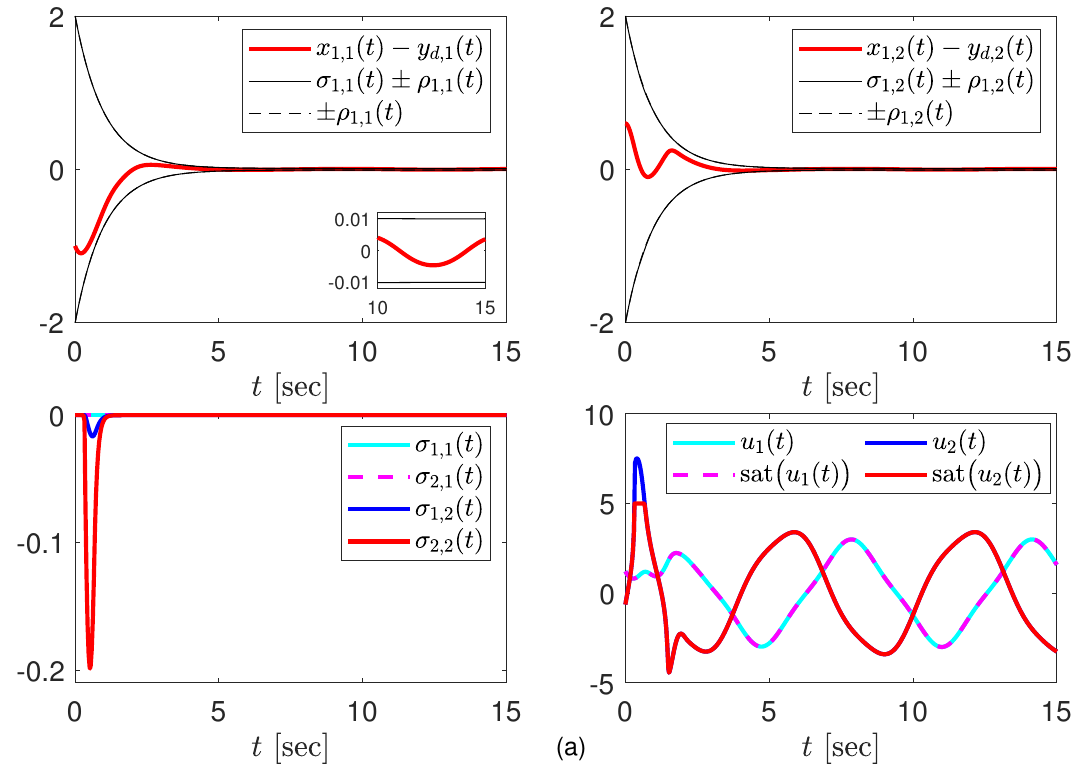}}\ \ \ \ 
	{\includegraphics[width=0.46\linewidth]{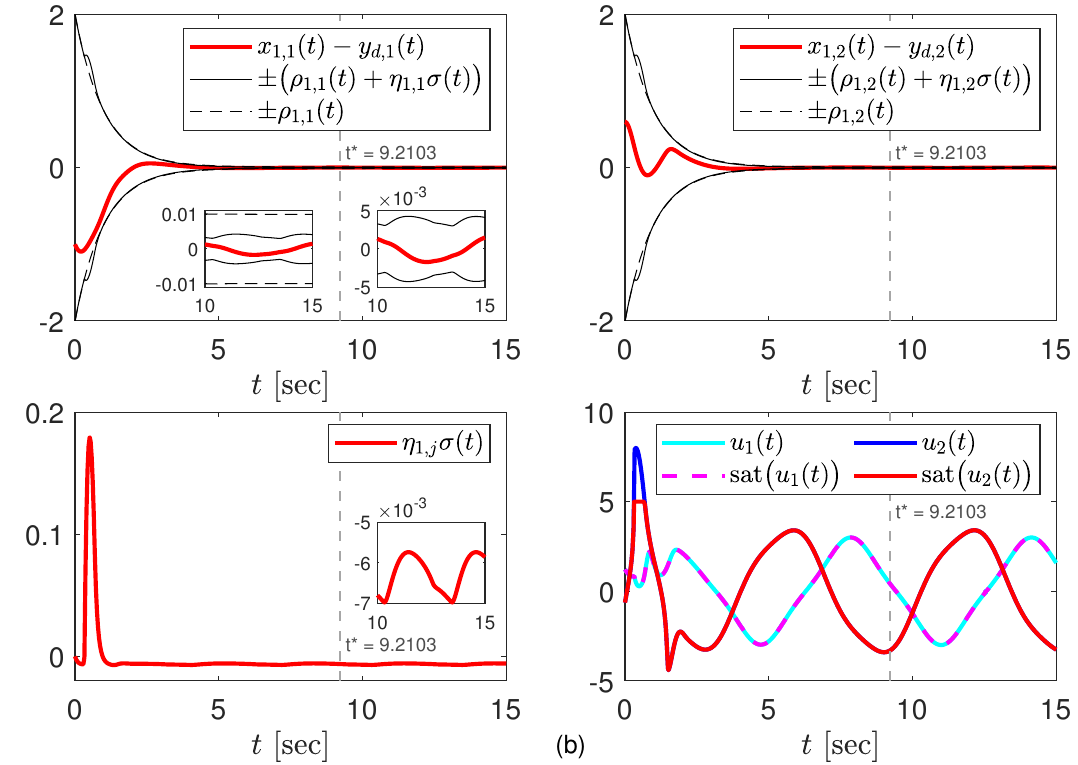}}
	\caption{%Comparative simulation experiments for output tracking control on the mass-spring-damper system (\ref{msd}): (a) simulation results of the control methodology developed in \cite{10273607}, where $\sigma_2(t)\in\mathbb R$ denotes the second-order modification signal; (b) simulation results of the control methodology developed in this work. From top to bottom and left to right, the output tracking performance constraints, the intermediate tracking performance constraints, the modification signals, the control input signal, and the saturated control input are depicted, respectively.
		Comparative simulation results: (a) method in \cite{10273607} where $\sigma_{2,j}(t)\in\mathbb R$ are the second-order modification signals; (b) method in this work.} \label{com}
\end{figure*}

To make the comparison fair, all settings for the simulation are equivalent to those in \cite{10273607}. Specifically, the initial conditions and saturation levels are:  $x_1(0)=[-1,1.6]^\top$, $x_{2,j}(0)=0$ and $\bar u_j=5$, $j\in\mathbb I_1^2$.
%Finally, the nonlinear mapping function is $T(\diamond)=\tan(\frac{\pi}{2}\diamond)$.
Moreover, the desired tracking trajectories are $y_{d}(t)=[\sin t,\cos t]^\top$. For the method in \cite{10273607}, the  design parameters for controller and modification signals, and the performance functions are: $k_{i,j}=1$, $\beta_{i,j}=10$ and $\rho_{i,j}(t)=(2.01-0.01)\exp^{-t}+0.01$. Note that the modified performance constraints in \cite{10273607} are 
\begin{flalign}
	|x_{1,j}(t)-\left( y_{d,j}(t)+\sigma_{1,j}(t) \right)|<\rho_{1,j}(t),\ \  \forall t\ge 0 \label{y}
\end{flalign}
where $\sigma_{1,j}(t)\in\mathbb R$ are the first-order  modification signals. To let (\ref{y}) consistent with the modified constraints (\ref{PC}), they are rewritten as
\begin{flalign}
	\sigma_{1,j}(t)-\rho_{1,j}(t)<e_j(t)<\sigma_{1,j}(t)+\rho_{1,j}(t),\ \   \forall t\ge 0.
\end{flalign}

For our method, the controller's and modification signal's design parameters $k_i$, $i\in\mathbb I_1^2$ and $\beta$ equal to those in \cite{10273607}. Next, we show that all assumptions and conditions for Theorem 1 are met. First, it can be readily verified that the numerical example is BIBS stable. Therefore, Assumption 1 naturally stands. Moreover, %from (\ref{f1}), 
we can trivially choose continuous functions $\bar f_1(x_1(t))=0$, $\bar f_2(\tilde x_2(t))=|-x_{2,1}(t)-2x_{1,1}(t)+x_{1,2}(t)-x_{1,1}^3(t)|+|-x_{2,2}(t)-2x_{1,2}(t)+x_{1,1}(t)-x_{1,2}^3(t)|+1$ and $\bar g_i(\tilde x_i(t))=2$  such that Assumption 2 is ensured. In addition, the selection of $y_d(t)$ makes Assumption 3 hold. Finally, we pick up the design parameters: $c_{1,j}=0.8$, $c_{2,j}=1$, $\gamma=0.02$, $\hat u=20$ and $\mu=4.1$, which makes the condition (\ref{c1}) stand.

The comparative simulation results are given in Fig. \ref{com}. It can be seen from Fig. \ref{com} that, under the effect of both control schemes, the output tracking control task for the example under input saturation and performance constraints are completed. However, the dynamics of the modification signals $\sigma_{i,j}(t)$ in \cite{10273607} is second-order and second dimension, which makes the structural and computational complexity of the controller in \cite{10273607} higher than ours. In addition, it can also be observed from Fig. \ref{com} (b) that, in our controller, the modification term $\eta_{1,j}\sigma(t)$ is less than zero at the \textcolor{black}{steady-state phase}, i.e., $\eta_{1,j}\sigma(t)\in[-0.007,-0.005]$ approximately, which thus leads to much better steady-state phase performance (i.e., $|x_{1,j}(t)-y_{d,j}(t)|<0.005$) than the one in \cite{10273607} (i.e., $|x_{1,j}(t)-y_{d,j}(t)|<0.01$). These facts %have confirmed 
confirm the superiority of our work. Finally, the method in \cite{10273607} is only applicable to MIMO systems with diagonal control gain matrices. In contrast, our control scheme is applicable to a broader class of highly-coupled MIMO systems with square gain matrices.

\section{Conclusion}
\label{sec:conclusion}
We studied the input-saturated %output tracking 
PPC problem for %a class of 
high-order uncertain highly-coupled MIMO  %strict-feedback 
nonlinear systems. %\textcolor{black}{To mitigate input saturation while alleviate control performance degradation,} 
To alleviate potential conflict while reducing conservatism of control performance in existing methods,  we constructed a bidirectional modification mechanism. Based on it, we proposed a low-complexity control scheme, ensuring that the modified performance constraints are satisfied and all closed-loop signals are bounded. A novel stability analysis framework was developed to bypass the obstacle in Lyapunov analysis. Future work includes the extension to non-BIBS stable or non-square MIMO nonlinear systems, and removal of parameter selection condition (\ref{c2}).

\section*{References}
\vspace{-1.5\baselineskip}
\bibliographystyle{ieeetr}
\bibliography{main}

\end{document}